\documentclass[
 reprint,
superscriptaddress,
 amsmath,amssymb,
 aps,
]{revtex4-2}

\usepackage[dvipdfmx]{graphicx}
\usepackage{dcolumn}
\usepackage{bm}
\usepackage{xcolor}
\usepackage{booktabs}
\usepackage[colorlinks=true, linkcolor=blue, citecolor=red, urlcolor=blue]{hyperref}
\newcommand{\dd}{{\rm d}}
\newcommand{\arcsec}{^{\prime\prime}}

\newcommand{\code}[1]{\texttt{#1}}

\begin{document}

\title{Microlensing constraints on Primordial Black Hole abundance \\
with Subaru Hyper Suprime-Cam  observations of Andromeda}

\author{Sunao Sugiyama}\email{sunaosugiyama@gmail.com}
\affiliation{Department of Physics and Astronomy, University of Pennsylvania, Philadelphia, PA 19104, USA}
\affiliation{Kavli Institute for the Physics and Mathematics of the Universe (WPI), 
The University of Tokyo Institutes for Advanced Study (UTIAS), 
The University of Tokyo, Chiba 277-8583, Japan}

\author{Masahiro Takada}
\affiliation{Kavli Institute for the Physics and Mathematics of the Universe (WPI), 
The University of Tokyo Institutes for Advanced Study (UTIAS), 
The University of Tokyo, Chiba 277-8583, Japan}
\affiliation{Center for Data-Driven Discovery (CD3), Kavli IPMU (WPI), UTIAS, 
The University of Tokyo, Kashiwa, Chiba 277-8583, Japan}

\author{Naoki Yasuda}
\affiliation{Kavli Institute for the Physics and Mathematics of the Universe (WPI), 
The University of Tokyo Institutes for Advanced Study (UTIAS), 
The University of Tokyo, Chiba 277-8583, Japan}
\affiliation{Center for Data-Driven Discovery (CD3), Kavli IPMU (WPI), UTIAS, 
The University of Tokyo, Kashiwa, Chiba 277-8583, Japan}

\author{Nozomu Tominaga}
\affiliation{National Astronomical Observatory of Japan, 2-21-1 Osawa, Mitaka, Tokyo 181-8588, Japan}
\affiliation{Astronomical Science Program, The Graduate University for Advanced Studies (SOKENDAI), 
2-21-1 Osawa, Mitaka, Tokyo 181-8588,
Japan}
\affiliation{Department of Physics, Konan University, 8-9-1 Okamoto, Kobe, Hyogo 658-8501, Japan}

\date{\today}

\begin{abstract}
We present updated microlensing analysis results based on high-cadence ($\sim$2~min) 
Subaru Hyper Suprime-Cam (HSC) observations of the Andromeda Galaxy (M31) in 2014, 2017, and 2020, 
yielding a total of 39.3 hours of data.
We use a point-lens 
finite-source 
model for the microlensing light curve model and employ
multi-stage selection procedures to identify microlensing candidates.
From more than 25,000 variable candidates detected across all nights, 
we identify 12 microlensing candidates with light-curve timescales shorter than 5~hours, 
and among them, 
4 secure candidates with high-significance detections.
We estimate detection efficiencies using light-curve-level simulations that 
account for observational conditions and 
finite-source 
effects.
Using a hierarchical Bayesian framework that combines the light-curve fitting information 
for each candidate with the Poisson statistics of the number of candidates, 
we derive constraints on parameters that characterize the abundance and mass scale of
primordial black hole (PBH) dark matter. 
First, we derive upper limits on the PBH abundance under the null hypothesis that
all events are assumed to be false detections.
Next, employing the PBH hypothesis in which all (or only secure) candidates
are assumed to be due to PBH microlensing, we derive the allowed region of the PBH parameters;
the inferred mass scale is $M_{\rm PBH}\sim10^{-7}$--$10^{-6}M_\odot$, and the PBH abundance
to the total dark matter is $f_{\rm PBH}\sim 10^{-1}$.
Our results demonstrate that HSC-M31 monitoring remains a uniquely powerful probe of PBHs, 
and highlight the need for further studies 
for example, using Rubin Observatory LSST observations of the Large Magellanic Cloud.
\end{abstract}

\maketitle

\section{Introduction}\label{sec:introduction}

Gravitational microlensing is a phenomenon in which the apparent brightness of 
a background star is temporarily amplified by the gravitational field of a 
compact object passing close to the line of sight to the star 
\citep{Paczynski.Paczynski.1986,Griest.Griest.1991}.
The characteristic timescale and magnification of a microlensing event are 
determined primarily by the mass of the lens and its relative transverse 
velocity with respect to the source and the observer.
Because microlensing does not rely on the luminosity of the lens, 
it provides a unique and powerful probe of otherwise invisible compact objects.

Microlensing surveys have been extensively used to detect and characterize 
exoplanets, and forthcoming wide-field, high-cadence surveys such as the 
Roman Space Telescope \citep{Spergel.Zhao.2015,Penny.Novati.2019} and 
the Vera C. Rubin Observatory (LSST) \citep{Ivezic.Zhan.2019,Collaboration.Zhan.2009} are 
expected to substantially advance our understanding of planetary populations 
through this technique.
Beyond exoplanet science, microlensing has also been widely applied to 
the study of non-luminous compact objects, including candidates for dark matter.

Primordial black holes (PBHs) are a particularly compelling dark matter candidate.
They may have formed in the early Universe from the collapse of primordial density 
fluctuations 
\citep{Zeldovich.Zeldovich.1972,Hawking.Hawking.1971,Carr.Hawking.1974,Khlopov.Khlopov.2007}
and could contribute a non-negligible fraction of the present-day ,
dark matter density 
\citep[e.g.,][]{Carr.Yokoyama.2020,Carr.Kuhnel.2021,Green.Kavanagh.2020}.
Constraining the PBH abundance, commonly parameterized by their fraction of 
the dark matter, requires statistically robust analyses of microlensing 
event rates across a wide range of timescales.

Early microlensing searches for compact dark matter objects were carried out by 
the MACHO and EROS collaborations,
establishing foundational methodologies for subsequent studies
stellar-mass range \citep[e.g.,][]{collaboration.Welch.2000,Tisserand.collaboration.2007}.
Subsequent long-timescale microlensing surveys by OGLE further tightened these
constraints using large samples of events toward the Galactic bulge
\citep{Wyrzykowski.Szewczyk.2010,Wyrzykowski.Rattenbury.2016}.
Later surveys, including OGLE \citep{Udalski.Szymanski.2015} and 
MOA \citep{Bond.Yock.2001}, 
have reported large samples of microlensing events, some of which exhibit 
light curves with timescales shorter than a day indicating isolated
low-mass lenses such as free-floating planets (FFPs)
\citep{Sumi.Ulaczyk.2011,Niikura.Masaki.2019}.
However, observations toward the Galactic bulge may suffer from significant 
contamination by FFPs, if the microlensing data are used to search for PBHs.

A very recent high-cadence microlensing experiment by the OGLE 
collaboration has provided updated constraints on primordial black holes 
in the sub-solar mass regime \citep{Mroz.Mroz.2024}.
In this study, two short-timescale candidate events were identified; 
however, detailed analyses suggested that one event is likely associated with 
a nearby dwarf star, while the other is consistent with a disk lens.
Based on the interpretation that no convincing PBH microlensing event was 
detected, the authors derived upper limits on the PBH dark matter fraction 
over the mass range from $M \sim 10^{-7} M_\odot$ to $1,M_\odot$, with the 
strongest constraint of $f_{\rm PBH} \lesssim 10^{-3}$ at $M \sim 10^{-6} M_\odot$.
These results demonstrate the strong potential of high-cadence 
microlensing surveys.
At the same time, it is useful to obtain complementary constraints 
using independent instruments, analysis pipelines, and different target 
fields and lines of sight, in order to further strengthen the overall 
picture of PBH abundance inferred from microlensing observations.

Observations of the Andromeda Galaxy (M31) offer a complementary and advantageous 
approach. The Subaru Telescope’s Hyper Suprime-Cam (HSC) 
\citep{Miyazaki.Yokota.2018,Komiyama.Wang.2018,Kawanomoto.Suzuki.2018,Furusawa.Lee.2018}
combines a large 8.2-meter aperture with excellent image quality and  
an exceptionally wide field of view, enabling simultaneous monitoring of a vast number of 
unresolved stars in M31 with high cadence.
In particular, the 1.5-degree diameter field of view of HSC closely matches 
the angular extent of M31, maximizing sensitivity to microlensing events caused 
by lenses lying in the Galactic and M31 halos.

Using a single night of HSC observations in 2014, \citet{Niikura.Chiba.2019} 
placed stringent constraints on the abundance of PBHs over a wide mass range, 
demonstrating the exceptional power of this dataset.
Despite the strength of these constraints, only one candidate microlensing 
event was identified, highlighting the importance of further analyses with 
expanded datasets and improved modeling techniques.

The HSC observations, characterized by short exposure times (90 seconds) and 
high cadence, are particularly well suited for detecting microlensing events caused 
by low-mass PBHs, which are expected to produce very short-duration signals.
Moreover, the halo-dominated line of sight toward M31 significantly 
reduces contamination from FFPs compared to Galactic bulge surveys, 
enhancing the reliability of PBH searches 
\citep{Crotts.Crotts.1992,Kerins.Collaboration.2001}.

Recent developments in microlensing analysis, especially the incorporation 
of finite-source effects \citep{Witt.Mao.1994,Gould.Gould.1994}, 
provide additional opportunities to improve sensitivity to short-timescale events.
While the standard Paczyński light curve assumes a point-like source, 
real stars have finite angular sizes, which can substantially modify 
the magnification profile, particularly for small Einstein radii or 
close lens–source alignments.
Finite-source modeling, typically implemented by convolving the point-source 
magnification with the surface brightness profile of the source star, 
enables more accurate characterization and selection of microlensing events.

In this paper, we present a comprehensive reanalysis of the 2014 HSC dataset 
and incorporate additional observations obtained in 2017 and 2020.
By expanding the temporal baseline and applying refined event selection and 
modeling techniques, we aim to improve constraints on the PBH abundance and 
to establish a robust framework for interpreting microlensing events in 
the context of dark matter studies.

This paper is organized as follows. 
In Section \ref{sec:data}, we describe the Subaru HSC observations of the 
Andromeda Galaxy and summarize the datasets used in this analysis. 
Section \ref{sec:data-analysis} details the data reduction procedures and 
the multi-stage event selection pipeline employed to identify microlensing candidates. 
In Section \ref{sec:efficiency}, we present our characterization of 
the detection efficiency based on light-curve-level simulations. 
Section \ref{sec:likelihood-model} introduces the likelihood framework and 
the hierarchical Bayesian methodology used to infer the parameters of 
the primordial black hole population. 
The resulting constraints on the primordial black hole abundance and 
mass scale are presented in Section \ref{sec:result}. 
Finally, we discuss the implications of our results and future prospects 
in Section \ref{sec:conclusion}.

\section{Data}\label{sec:data}
\begin{table}
    \centering
    \caption{Summary of the HSC datasets used in this paper. 
    The first column lists
    the observing date.
    The second column denotes the PI of the observation.
    The third column denotes
    the number of allocated nights for each run.
    The fourth column denotes
    the filters used in each observation.
    The fifth column denotes
    the number of exposures
    taken on each observing
    date. The sixth column, denotes as
    $T_{\rm eff}$~(in units of hours), indicates
    the effective duration of the observation.
    Here the duration accounts for the period 
    when the observation was halted due to
    focusing or
    bad weather conditions such as high humidity.
    The last column, denoted as ``PSF'',
    indicates the typical seeing size.}
    \begin{tabular}{ccccccc}
        \hline\hline
        Date & PI & Nights & Filter & Images & $T_{\rm eff}$ & PSF
        \\
        \hline
        2014-11-24 & Takada   & 1.0 & $r$  & 189 & 6.4 & $0.68\arcsec$ \\
        2017-09-20 & Takada   & 1.0 & $r2$ & 215 & 7.1 & $0.86\arcsec$ \\
        2020-10-21 & Sugiyama & 0.5 & $r2$ & 152 & 5.0 & $0.80\arcsec$ \\
        2020-10-22 & Sugiyama & 0.5 & $r2$ & 156 & 5.2 & $0.84\arcsec$ \\
        2020-11-11 & Sugiyama & 0.5 & $r2$ &  56 & 1.8 & $1.13\arcsec$ \\
        2020-11-12 & Sugiyama & 0.5 & $r2$ & 126 & 4.1 & $1.38\arcsec$ \\
        2020-11-14 & Sugiyama & 0.5 & $r2$ & 152 & 5.1 & $0.80\arcsec$ \\
        2020-11-16 & Sugiyama & 0.5 & $r2$ &   0 & 0.0 & -- \\
        2020-11-18 & Sugiyama & 0.5 & $r2$ &  34 & 0.0 & -- \\
        2020-11-20 & Sugiyama & 0.5 & $r2$ & 140 & 4.6 & $1.08\arcsec$ \\
        \hline\hline
    \end{tabular}
    \label{tab:obs-summary}
\end{table}

We carried out a series of monitoring observations of stars in 
Andromeda Galaxy (M31), as summarized in Table~\ref{tab:obs-summary}.
The first observation was carried out on November 24, 2014.
One night was allocated for this observation and 189 images were
obtained, each of which has an exposure of 90 seconds 
and a read out of 30 seconds, resulting in 2-minute cadence.
Excluding data affected by bad weather and other overheads, 
the effective science-quality observing time 
for this night amounts to a total of 6.4~hours.
The typical seeing size is $0.68\arcsec$.
The $r$-band filter was replaced with a
new one ($r2$) on
July 28, 2016 (HST), and all our observations after 2017-09-20 were 
conducted using the $r2$ filter~
\footnote{\url{https://subarutelescope.org/Instruments/HSC/sensitivity.html}}.

The second observation was carried out on September 20, 2017.
We obtained 215~images with a total effective observing time 
of $T_{\rm eff}=7.1$~hours, under 
typical seeing conditions of $0.86\arcsec$.

The other eight observations were carried out in 2020. 
In this set of observations, 8~half-nights were allocated,
yielding a net total of 4~full nights. 
We obtained science-quality data on November 12, 14 and 20 and 
partially on November 11, while the data collected on 
November 16 and 18 did not meet the expected quality
due to poor weather conditions.
In total, we use 39.3~hours of HSC data in this paper.

\section{Data analysis}\label{sec:data-analysis}
\subsection{Standard image processing}\label{ssec:standard-image-processing}

As a first step of the data analysis, we performed the standard image 
processing using the HSC pipeline~\citep{Bosch.Yamanoi.2018}, 
which is a dedicated data reduction pipeline for HSC imaging data. 
We used the latest version \code{hscpipe8} in this paper,
while \citet{Niikura.Chiba.2019} used \code{hscpipe3}. 
The standard image processing includes a series of image calibrations; 
bias subtraction, flat fielding with dome flats, coadding, astrometry 
and photometry calibrations. 
We refer the readers to \citet{Aihara.Yamashita.2021} for detail of \code{hscpipe8}.

\subsection{Image subtraction and object detection}
\label{ssec:image-subtraction}
\begin{table}
    \centering
    \caption{Summary of the master catalog of variable-star candidates.}
    \begin{tabular}{ccc}
        \hline\hline
        Name of Master catalog & Number of events
        \\
        \hline
        2014-11-24 & 7281 & \\
        2017-09-20 & 6199 & \\
        2020-10-21 & 3636 & \\
        2020-10-22 & 1582 & \\
        2020-11-11 & 348  & \\
        2020-11-12 & 2029 & \\
        2020-11-14 & 2504 & \\
        2020-11-16 & 0    & \\
        2020-11-18 & 0    & \\
        2020-11-20 & 1714 & \\
        \hline\hline
    \end{tabular}
    \label{tab:catalog}
\end{table}
\begin{figure*}
\includegraphics[width=\textwidth]{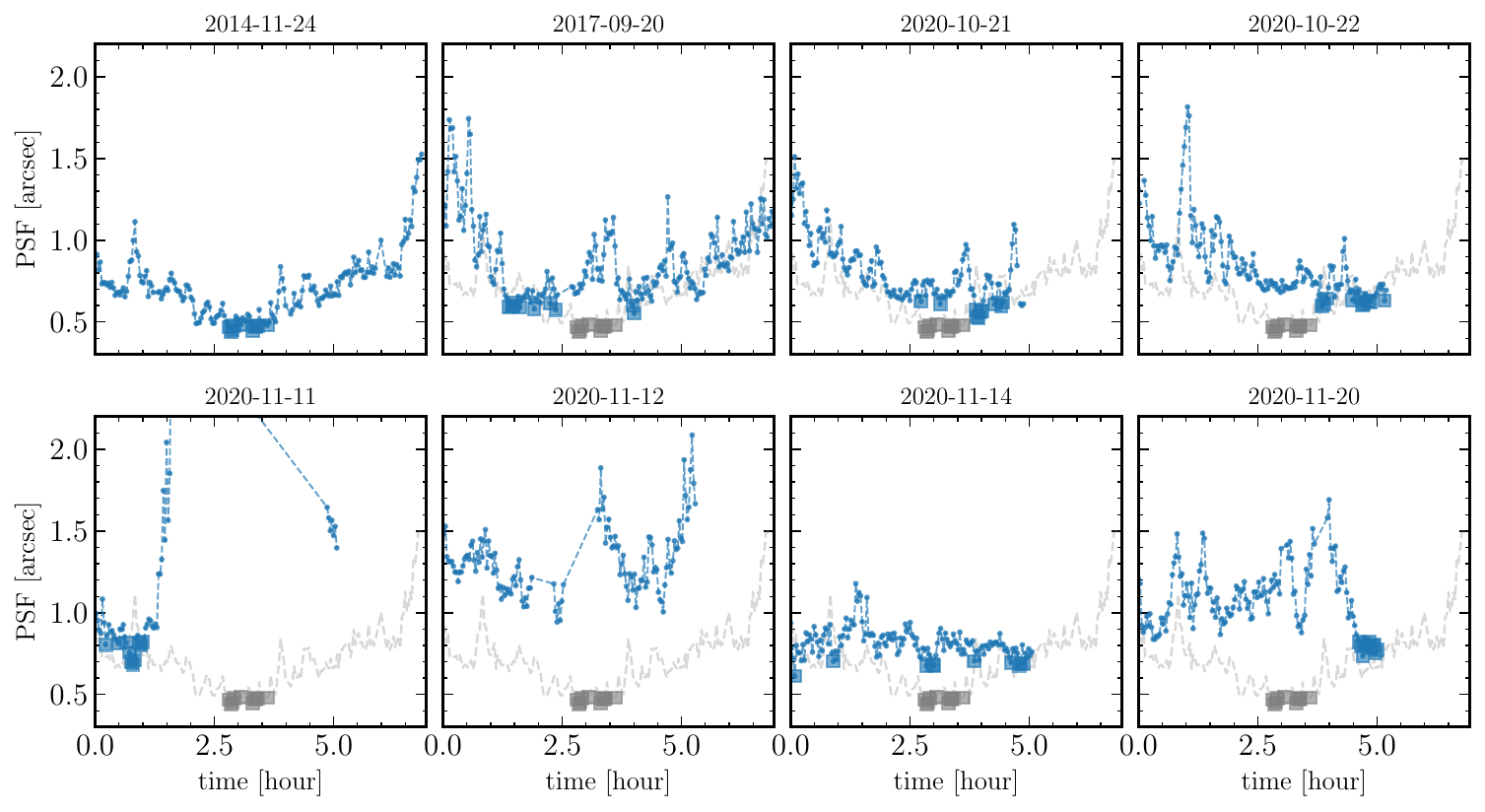}
    \caption{
    In each panel, the solid line indicates the PSF size (seeing) 
    as a function of the observation time from the beginning of 
    observation on each night, and the square points 
    indicate the 10 images 
    with the best seeing on each night, which are therefore used
    to generate the reference images for image subtraction, as described 
    in Section~\ref{ssec:image-subtraction}. 
    The PSF size of the 2014-11-24 data is overplotted in gray 
    in each panel for comparison. Note that the 2020-11-12 data 
    have relatively large seeing 
    compared to other observation nights, and 
    therefore we do not generate a reference image for this night.
    We also note that, although the 2020-11-14 data are under 
    good conditions throughout the night, 
    the reference image was generated from images with relatively 
    large seeing compared to those from the 2014-11-24 and 2017-09-20 data, 
    which are the best data in our dataset.
    }
    \label{fig:seeing}
\end{figure*}

Since M31 is a dense stellar field and our microlensing analysis lies 
in the pixel lensing regime, we need to employ the difference image technique 
to detect variable candidates in the images \citep{Gould.Gould.1995}. 
The difference image is developed 
in \citet{Alard.Lupton.1997,Alard.Alard.2000}, and has been implemented in \code{hscpipe8}.

As we are interested in microlensing events on timescales within a single night, 
we performed the image subtraction for the HSC data taken 
during each night listed in Table~\ref{tab:obs-summary}.
First, we co-added the 10 images with the best seeing conditions
to create a reference image for each night. 
Fig.~\ref{fig:seeing} shows the PSF size as a function of the 
elapsed time since the start of the observations for each night. 
The seeing was good in the 2014-11-24 data, and the 10 images 
used to generate the reference image have
a mean seeing size of $0.47\arcsec$. 
For the 2017-09-20, 2020-10-21, and 2020-10-22 data, the seeing was
overall good, and the 10 images used to generate the reference image have 
mean seeing sizes of $0.59\arcsec$--$0.62\arcsec$
for each night. For 
the 2020-11-14 data, the seeing was stable throughout the observing hours,
and the 10 images used for the reference image have a mean seeing size of 
$0.68\arcsec$, which is worse than those on the aforementioned dates.  
For the 2020-11-11, 2020-11-12 and 2020-11-20 data, 
the seeing was poor, and the data are not usable for the microlensing analysis, 
as we will show below.
In particular we were not able to generate a useful reference image 
for the 2020-11-12 data.

Next, we co-added three consecutive images to generate a target image 
at each epoch.
Finally, we subtracted the reference image from each target image to 
produce a difference image, from which we detected variable object candidates.
Here we defined variable candidates as local minima or maxima in 
each of the difference images. 
To obtain a reliable catalog of variable candidates, we applied
the following selection criteria:
\begin{itemize}
    \item PSF magnitude cut: A candidate must have $5\sigma$ or higher PSF 
    magnitude in at least two difference images. The candidates in 
    different difference images are identified as the same object when 
    their positions is within 2~pixels of each other.
    \item Size cut: The size of a candidate should be consistent with 
    the PSF size within 25\%.
    \item Roundness cut: Candidates should have a round shape, requiring 
    the axis ratio of the object to be greater than $0.75$, as 
    the HSC PSF is nearly round.
    \item Shape cut: Candidates are required to have a shape consistent 
    with PSF; specifically, the residual image, obtained by subtracting 
    a rescaled PSF from the difference image, must have 
    a cumulative residual smaller than $3\sigma$.
\end{itemize}
These selection criteria yield reliable candidate objects, 
which we refer to as the 
master catalog for the data of each observation date.
The number of the objects in each master catalog are summarized 
in Table~\ref{tab:catalog}.

Once we obtained the master catalog for a given observation night,
we measured the forced photometry for each candidate on each exposure 
image of that night, at the position where the object was detected. 
For the photometry error at each epoch, we use the variance of 
the photometry measurements at 1,000 random 
points in the corresponding patch,  following the method in 
Ref.~\cite{Niikura.Chiba.2019}. 
The error estimated in this way includes the large-scale 
residual background arising from imperfect background subtraction.

\subsection{Event Selection}\label{ssec:event-selection}
\begin{table*}
    \setlength{\tabcolsep}{12pt}
    \centering
    \caption{
    Summary statistics of each candidate for selection.
    See Section~\ref{ssec:event-selection} for the details.}
    \begin{tabular*}{\textwidth}{cl}
        \hline\hline
        Name      & Definition \\
        \hline
        \code{bump}    & 
        The time interval during which
        $\Delta F_i - \overline{\Delta F}>2\sigma(\Delta F_i)$,
        being satisfied consecutively over a time interval \\
        \code{bumplen} & The duration or the number of images of the longest bump \\
        $t_{\rm FHWM}$ & The FWHM time interval computed from the best-fit microlensing light-curve model \\
        \code{mlc2}    & The chi-square value of the best-fit microlensing light-curve model, integrated over the fitting time interval \\
        \code{mlc2i}   & Same as \code{mlc2}, but using only the images with $|t_i-t_0|<t_{\rm FWHM}$\\
        \code{mlc2o}   & Same as \code{mlc2}, but using only the images with $|t_i-t_0|>t_{\rm FWHM}$\\
        \code{mlc2il}  & Same as \code{mlc2i}, but using only the images before the peak time \\
        \code{mlc2ir}  & Same as \code{mlc2i}, but using only the images after  the peak time \\
        \code{mlc2ol}  & Same as \code{mlc2o}, but using only the images before the peak time \\
        \code{mlc2or}  & Same as \code{mlc2o}, but using only the images after  the peak time \\
        \code{asym}    & Asymmetry parameter of the light curve, as defined in Eq.~(\ref{eq:asym}) \\
        \code{scorr}   & The cross-correlation between $\Delta F$ and the seeing size at each epoch, as defined in Eq.~(\ref{eq:scorr})\\
        \code{scorrs}  & The cross-correlation between $\Delta F$ and the seeing size for a certain time interval (see text for the details)\\
        $\max{A}$      & The maximum magnification predicted by the best-fit microlensing model\\
        \code{mlsig}   & The significance of the peak detection as a micro-lensing light-curve event, as defined in Eq.~(\ref{eq:mlsig})\\
        \hline\hline
    \end{tabular*}
    \label{tab:selection-statistics}
\end{table*}

\begin{table*}
    \setlength{\tabcolsep}{5pt}
    \centering
    \caption{Selection criteria for microlensing candidates.
    See Section~\ref{ssec:event-selection} for the details.}
    \begin{tabular*}{\textwidth}{cccl}
        \hline\hline
        Level & Label & Criterion & Purpose \\
        \hline
        1 & bump   & $\code{bumplen}\geq 3.0$ & 
        a light-curve peak spanning at least three consecutive detected data points \\
        1 & mlc2   & $\code{mlc2}\leq 3.0$    & 
        a reasonable goodness-of-fit to a microlensing model\\
        1 & mlc2i  & $\code{mlc2i}\leq3.0$   & 
        a reasonable goodness-of-fit around the peak \\
        1 & mlc2o  & $\code{mlc2o}\leq3.0$  & 
        a reasonable goodness-of-fit away from the peak \\
        1 & mlc2il & $\code{mlc2il}\leq3.0$   & 
        a reasonable goodness-of-fit on the left side of the peak \\
        1 & mlc2ir & $\code{mlc2ir}\leq3.0$   & 
        a reasonable goodness-of-fit on the right side of the peak \\
        1 & mlc2ol & $\code{mlc2ol}\leq3.0$  & 
        a reasonable goodness-of-fit on the left side, away from the peak\\
        1 & mlc2or & $\code{mlc2or}\leq3.0$  & 
        a reasonable goodness-of-fit on the right side,
        away from the peak \\
        1 & asym   & $\code{asym}<0.17$       & 
        a light curve symmetric around the peak \\
        1 & scorr  & $\code{scorr}/\sqrt{\rm dof}\leq3.0$ & 
        no significant correlation with the seeing\\
        1 & scorrs & $\max\left[\code{scorrs}/\sqrt{\rm dof}\right]\leq3.0$ & 
        no significant correlation with the seeing 
        on a certain timescale\\
        1 & ntscale & $t_{\rm FWHM}/t_{\rm obs}\leq1$ & 
        the light-curve timescale detectable within the observing window\\
        1 & amax   & $\max{A}\geq1.34$        & 
        a significant peak magnification \\
        1 & mlsig  & $\code{mlsig}\geq5$      & 
        a significant detection of the light curve integrated over the observing window\\
        2 & mlvs   & visual inspection        & remove impostors due to moving objects or imperfect
        image subtraction \\
        3 & mlnrpt & no repeatability          & 
        no repeated variation at the candidate position on other observing nights \\
        \hline\hline
    \end{tabular*}
    \label{tab:selection-criteria}
\end{table*}

\begin{table}
    \setlength{\tabcolsep}{8pt}
    \centering
    \caption{
    Parameters and prior ranges
    of point-lens finite-source (PLFS) microlensing light-curve model, 
    used to fit each candidate's light curve (see text for the details); 
    $f_0$ is a parameter to model the intrinsic flux of a source star;
    $u_0$ is the impact parameter at closest approach 
    of the lens to the source on the sky; $t_0$ is the time of peak magnification;
    $t_{\rm E}$ is the Einstein microlensing timescale; $\rho$ is the source radius
    in units of the Einstein radius.
    $t_{0}^\code{bump}$ denotes the estimated time of each candidate's light-curve peak, 
    and $\sigma(t_{0}^\code{bump})$ is the $1\sigma$ uncertainty.
    The lower bound of the $t_{\rm E}$ prior is set to 2~minutes, 
    which is the shortest timescale in our cadence. 
    The upper bound is set to $t_{\rm obs}$, the duration of 
    the observing window for each night.  
    }
    \begin{tabular}{cl}
    \hline\hline
    Parameter & Prior \\
    \hline
    $f_0$ & flat$[-0.5, 0.5]$ \\
    $u_0$ & flat$[0.0, 1.5]$ \\
    $t_0$ & flat$[t_{0}^\code{bump}-\sigma(t_{0}^\code{bump}), t_{0}^\code{bump}+\sigma(t_{0}^\code{bump})]$ \\
    $t_{\rm E}$ & flat$[2{\rm min}, 
    t_{\rm obs}]$ \\
    $\rho$ & flat$[0.0, 1.5]$\\
    \hline\hline
    \end{tabular}
    \label{tab:plfs-prior}
\end{table}
\begin{table*}
    \centering
    \caption{
    The number of the microlensing candidates that passed each of 
    the selection criteria listed in Table~\ref{tab:selection-criteria}.}
\begin{tabular}{llllllllllllllllll}
\toprule
        ID & master & bump & mlc2 & mlc2i & mlc2o & mlc2il & mlc2ir & mlc2ol & mlc2or & asym & scorr & scorrs & ntscale & amax & mlsig & mlvs & mlnrpt \\
\midrule
2014-11-24 &   7281 & 5139 & 1135 &   938 &   911 &    878 &    819 &    814 &    783 &  391 &    72 &     70 &      57 &   53 &    38 &   25 &      5 \\
2017-09-20 &   6199 & 4554 & 1060 &   962 &   907 &    877 &    837 &    821 &    800 &  287 &   142 &     59 &      50 &   44 &    18 &   18 &      7 \\
2020-10-21 &   3636 & 3348 &  361 &   218 &   206 &    205 &    196 &    196 &    195 &   13 &     5 &      5 &       3 &    3 &     0 &    0 &      0 \\
2020-10-22 &   1582 & 1561 &   66 &    52 &    47 &     44 &     41 &     40 &     38 &   18 &     1 &      1 &       1 &    1 &     0 &    0 &      0 \\
2020-11-11 &    348 &  274 &   16 &    16 &    15 &     12 &     11 &     11 &     11 &    2 &     0 &      0 &       0 &    0 &     0 &    0 &      0 \\
2020-11-12 &   2029 & 1581 &  202 &   185 &   185 &    180 &    154 &    154 &    154 &   23 &    18 &     18 &       0 &    0 &     0 &    0 &      0 \\
2020-11-14 &   2446 & 1959 &  359 &   343 &   326 &    312 &    278 &    270 &    261 &   38 &    23 &     22 &      18 &   17 &     8 &    7 &      0 \\
2020-11-20 &   1714 & 1241 &  260 &   223 &   215 &    205 &    165 &    164 &    164 &   15 &     4 &      4 &       1 &    1 &     0 &    0 &      0 \\
\bottomrule
\end{tabular}
    \label{tab:selection-result}
\end{table*}
\begin{table*}
    \centering
    \caption{
    Summary of the 12 microlensing candidates obtained 
    after applying the selection criteria to the master catalog.
    Each candidate is identified by the combination of the
    master catalog, patch id, and objid.
    The second group of columns summarizes the RA, DEC position of the candidates.
    The third group of columns summarizes some of the summary statistics 
    used for the candidate selection, with the degrees of freedom 
    that are used to compute these summary statistics.
    The last column group summarizes the best-fit parameters of 
    the PLFS microlensing model. 
    Note that the parameter degeneracies are generally strong, and 
    the parameter values shown here represent
    a nominal set of the parameters obtained from the Monte-Carlo chains 
    of the light-curve fitting. 
    The secure candidates are marked by $\dagger$, 
    and are defined as those that further satisfy $\code{mlsig}/{\rm dof}>10$.
    }
    \resizebox{\textwidth}{!}{
\begin{tabular}{@{}cl@{\hspace{1.6em}}ll@{\hspace{1.6em}}rrrrrr@{\hspace{1.6em}}rrrrr@{}}
\toprule
 & ID & \multicolumn{2}{c}{Coord.} & \multicolumn{6}{c}{Selection} & \multicolumn{5}{c}{ML params} \\
 & master/patch/objid & RA & Dec & mlc2 & asym & scorr & $\max A$ & mlsig & dof & $f_0$ & $u_0$ & $t_0$ & $t_{\rm E}$ & $\rho$ \\
\midrule
$\dagger$ & 2014-11-24/1,5/75 & $00^{\mathrm{h}}\mkern1mu46^{\mathrm{m}}\mkern1mu38.41^{\mathrm{s}}$ & $+41^{\circ}\mkern1mu16^{\prime}\mkern1mu43.2^{\prime\prime}$ & 232.68 & 0.08 & 0.07 & 2.88 & 2486.89 & 189 & -0.43 & 0.21 & 16613.35 & 5598.13 & 0.73 \\
$\dagger$ & 2014-11-24/4,2/63 & $00^{\mathrm{h}}\mkern1mu42^{\mathrm{m}}\mkern1mu53.02^{\mathrm{s}}$ & $+40^{\circ}\mkern1mu48^{\prime}\mkern1mu07.6^{\prime\prime}$ & 251.56 & 0.08 & -0.03 & 1.90 & 3085.37 & 189 & -0.50 & 0.85 & 16430.47 & 8654.20 & 0.93 \\
 & 2014-11-24/6,7/168 & $00^{\mathrm{h}}\mkern1mu41^{\mathrm{m}}\mkern1mu20.91^{\mathrm{s}}$ & $+41^{\circ}\mkern1mu41^{\prime}\mkern1mu16.1^{\prime\prime}$ & 120.54 & 0.15 & -0.06 & 6.10 & 1453.65 & 189 & -0.33 & 0.05 & 16172.20 & 10218.13 & 0.33 \\
 & 2014-11-24/6,7/169 & $00^{\mathrm{h}}\mkern1mu41^{\mathrm{m}}\mkern1mu07.94^{\mathrm{s}}$ & $+41^{\circ}\mkern1mu48^{\prime}\mkern1mu54.0^{\prime\prime}$ & 147.72 & 0.12 & -0.06 & 2.21 & 1281.99 & 189 & -0.27 & 0.68 & 15587.25 & 5763.05 & 0.83 \\
 & 2014-11-24/7,6/96 & $00^{\mathrm{h}}\mkern1mu40^{\mathrm{m}}\mkern1mu07.90^{\mathrm{s}}$ & $+41^{\circ}\mkern1mu36^{\prime}\mkern1mu17.8^{\prime\prime}$ & 120.79 & 0.13 & 0.14 & 2.56 & 1335.64 & 189 & -0.32 & 0.48 & 16223.85 & 5374.24 & 0.77 \\
 & 2017-09-20/1,4/46 & $00^{\mathrm{h}}\mkern1mu46^{\mathrm{m}}\mkern1mu21.70^{\mathrm{s}}$ & $+41^{\circ}\mkern1mu09^{\prime}\mkern1mu17.1^{\prime\prime}$ & 255.28 & 0.14 & 0.10 & 1.52 & 1092.90 & 214 & -0.45 & 0.87 & 16060.40 & 3632.56 & 0.52 \\
 & 2017-09-20/3,4/31 & $00^{\mathrm{h}}\mkern1mu44^{\mathrm{m}}\mkern1mu07.74^{\mathrm{s}}$ & $+41^{\circ}\mkern1mu05^{\prime}\mkern1mu17.6^{\prime\prime}$ & 107.61 & 0.05 & -0.00 & 3.75 & 1488.64 & 214 & -0.50 & 0.31 & 14589.27 & 16286.60 & 0.25 \\
 & 2017-09-20/4,2/9 & $00^{\mathrm{h}}\mkern1mu42^{\mathrm{m}}\mkern1mu56.85^{\mathrm{s}}$ & $+40^{\circ}\mkern1mu49^{\prime}\mkern1mu09.8^{\prime\prime}$ & 334.17 & 0.07 & -0.13 & 2.75 & 1461.36 & 214 & -0.50 & 0.46 & 4088.65 & 4814.50 & 0.42 \\
$\dagger$ & 2017-09-20/5,7/45 & $00^{\mathrm{h}}\mkern1mu42^{\mathrm{m}}\mkern1mu09.81^{\mathrm{s}}$ & $+41^{\circ}\mkern1mu44^{\prime}\mkern1mu38.3^{\prime\prime}$ & 213.67 & 0.07 & -0.08 & 2.06 & 2210.57 & 213 & -0.41 & 0.74 & 14173.11 & 7451.41 & 0.92 \\
 & 2017-09-20/7,4/36 & $00^{\mathrm{h}}\mkern1mu39^{\mathrm{m}}\mkern1mu47.83^{\mathrm{s}}$ & $+41^{\circ}\mkern1mu11^{\prime}\mkern1mu05.0^{\prime\prime}$ & 140.92 & 0.12 & -0.09 & 3.00 & 1121.77 & 214 & -0.45 & 0.47 & 14942.25 & 10821.83 & 0.57 \\
 & 2017-09-20/7,6/26 & $00^{\mathrm{h}}\mkern1mu40^{\mathrm{m}}\mkern1mu13.54^{\mathrm{s}}$ & $+41^{\circ}\mkern1mu31^{\prime}\mkern1mu29.5^{\prime\prime}$ & 184.09 & 0.10 & -0.07 & 3.44 & 1728.40 & 214 & -0.43 & 0.39 & 15833.30 & 12398.83 & 0.40 \\
$\dagger$ & 2017-09-20/8,2/119 & $00^{\mathrm{h}}\mkern1mu39^{\mathrm{m}}\mkern1mu33.27^{\mathrm{s}}$ & $+40^{\circ}\mkern1mu50^{\prime}\mkern1mu06.6^{\prime\prime}$ & 107.61 & 0.15 & -0.14 & 2.10 & 2467.61 & 125 & -0.22 & 0.74 & 15436.09 & 9623.68 & 0.78 \\
\bottomrule
\end{tabular}
    }
    \label{tab:candidate-catalog}
\end{table*}

Our selection process of microlensing events consists of three levels. 
The first-level selection is the light-curve based selection, 
where the selection is applied to the derived summary statistics of 
each light curve. 
The second-level selection is based on the images to visually 
assess if the events are really likely due to microlensing 
or merely imposters. 
The third-level selection is based on the light curves and 
images taken in other observing dates at the detection sky position. 
This ensures that the selected events are not repeating variables.

We start with the first-level selection. 
In what follows, we denote the flux in the difference image as $\Delta F_i$, 
where the subscript $i$ refers to the $i$-th image in the light curve 
of a given candidate.

Since the light curve of a microlensing event has a peak at the time 
of the closest approach between the lens and the source on the sky, 
we first select candidates whose light curves exhibit a peak.
For this purpose, we define a bump in each light curve as 
a time interval during which the difference flux is consecutively 
above the mean difference flux
\begin{align}
    \Delta F_i - \overline{\Delta F}>2\sigma(\Delta F_i).
\end{align}
Here, the mean difference flux $\overline{\Delta F}$ is defined 
as the average over all the time indices. 
Note that a single light curve may have multiple bumps.
Then we define the length of a bump in a light curve, 
denoted as \code{bumplen}, by the number of images within the bump. 
If a light curve has multiple bumps, we define 
\code{bumplen} as the largest \code{bumplen} among them.
If a light-curve has no bump, we set $\code{bumplen}=0$. 
The first selection criterion is based on the bump length, requiring
$\code{bumplen}\geq3$.
For the later use, we also define the most significant bump 
as the one with the highest bump significance:
\begin{align}
    \code{bumpsignif} = \sum_{i\in \code{bump}} \delta\Delta F_i.
\end{align}
Using the start and end times for the most significant bump, 
denoted as $t_{\rm start}^\code{bump}$ and $t_{\rm end}^\code{bump}$,
we define a rough estimate of the peak time as 
$t_{0,\code{bump}} = (t_{\rm end}^\code{bump}+t_{\rm start}^ \code{bump})/2$ 
and its uncertainty as
$\sigma(t_{0}^\code{bump}) = (t_{\rm end}^\code{bump}-t_{\rm start}^\code{bump})/2 + \Delta t$,  
where $\Delta t=2~\text{minutes}$ is the minimum time resolution, 
i.e. the cadence of our observations.

The second step is to fit the microlensing light-curve model to 
the measured light curve. With a model of the magnification of 
the microlensing $A(u)$, we model the difference flux as 
\begin{align}
    \Delta F_i^{\rm model} &= F_0 \Delta A(u_i) \equiv F_0\left[A(u_i) - \frac{1}{10}\sum_{i'\in {\rm ref}} A(u_{i'})\right] \label{eq:df-model}
\end{align}
where
\begin{align}
    u_i &= \sqrt{u_0^2 + \left(\frac{t_i-t_0}{t_{\rm E}}\right)^2}.
\end{align}
Here, we define the intrinsic flux $F_0$, the peak time $t_0$ 
(the closest approach time), the impact parameter $u_0$, and 
the Einstein time $t_{\rm E}$. The second term in Eq.~(\ref{eq:df-model}) 
corresponds to the flux of the reference image that was constructed by 
co-adding the images of 10 best-seeing epochs
\footnote{Note that in \citet{Niikura.Chiba.2019} we used the 
representative time for the reference image $t_{\rm ref}$ which is 
defined by the mean time.}.

For the model of the microlensing magnification $A(u)$, we use the 
point-lens finite-source (PLFS) model:
\begin{align}
    A_{\rm PLFS}(u; \rho) = \int{\rm d}\bm{r} s(\bm{r}; \rho) A_{\rm PLPS}(\bm{u}+\bm{r})
    \label{eq:APLFS}
\end{align}
where $A_{\rm PLPS}(u)=(u^2+2)/u/\sqrt{u^2+4}$ is the point-lens 
point-source magnification, also known as Paczyński light curve 
\citep{Paczynski.Paczynski.1986}. 
To account for the finite source size effect \citep{Witt.Mao.1994}, 
we assume a circular 
disk profile with uniform brightness, given by
$s(u; \rho)=\Theta(u-\rho)/2\pi\rho^2$, where
$\Theta$ is the step function. The finite-source parameter $\rho$ is 
the angular size of the source star normalized by the angular Einstein radius, 
$\rho=\theta_{\rm s}/\theta_{\rm E} = (R_{\rm s}/d_{\rm s})/ (R_{\rm E}/d_{\rm l})$. 
We use the public code \code{fastlens}
\footnote{\url{https://github.com/git-sunao/fft-extended-source}} 
developed in \citep{Sugiyama.Sugiyama.2022} to perform the integral 
in Eq.~(\ref{eq:APLFS}). Around each time step $t_i$, we average the 
difference flux model over an interval of $\pm\Delta t/2$ to account 
for the finite exposure time.
For the search of best-fit parameters, we use a nested sampling rather 
than a minimizer to avoid 
getting trapped
into local minima of the parameter space. As the sampling process proceeds, 
we increase the temperature of the likelihood to speed up the inference.

Eqs.~(\ref{eq:df-model})--(\ref{eq:APLFS}) define
the model of the microlensing light-curve, which is characterized by
five model parameters $\bm{\theta}=\{F_0, u_0, t_0, t_{\rm E}, \rho\}$. 
We estimate the best-fit parameters by minimizing the chi-square
\begin{align}
    \chi^2(\bm{\theta}) = \sum_i \left(\frac{\Delta F_i - \Delta F_i^{\rm model}}{\sigma(\Delta F_i)}\right)^2 .
    \label{eq:chi-square}
\end{align}
For efficient parameter inference, we use priors on model parameters, 
as summarized in Table~\ref{tab:plfs-prior}. 
In the PLFS model, the light curve has a single maximum,
so we can infer the intrinsic flux by comparing the maximum difference flux 
$\max(\Delta F)$ with the maximum of the difference magnification $\max(\Delta A)$. 
To ensure sufficient flexibility, we allow the intrinsic flux parameter to vary 
by up to 50\% from the estimated value: to be precise, we introduce the fractional parameter 
$f_0\in[-0.5, 0.5]$ to define $F_0 = \max(\Delta F_i)/\max(\Delta A) (1+f_0)$ 
\cite{Niikura.Chiba.2019}, 
and use $f_0$ as a free parameter instead of $F_0$.
To summarize, our model has five parameters $\bm{\theta}=\{f_0, u_0, t_0, t_{\rm E}, \rho\}$.

Using the best-fit parameters of the PLFS model obtained from 
the light-curve fitting, we compute the derived statistics 
for selection of microlensing candidates. 
The first one is the full-width half-maximum (FWHM) time of the magnification:
\begin{align}
    A(t_0 + t_{\rm FWHM}) = 1+\frac{\max{A}-1}{2}.
    \label{eq:fwhm-time}
\end{align}
In PLFS model, the FWHM time is a function of $\{u_0, t_{\rm E}, \rho\}$. 
We apply a selection criterion on the FWHM timescale, $t_{\rm FWHM}/t_{\rm obs}\leq1$
to exclude events with timescales longer than the observation duration.
We define the global goodness-of-fit by the chi-square in Eq.~(\ref{eq:chi-square}) 
at the best fit model, denoted as \code{mlc2}.
We also define the chi-square values, \code{mlc2i} and \code{mlc2o}, 
using only the data points in ``in'' and ``out'' time regime, defined by 
$|t_i-t_0|<t_{\rm FWHM}$ and $|t_i-t_0|>t_{\rm FWHM}$, respectively. 
Similarly, we define \code{mlc2il}, \code{mlc2ir}, \code{mlc2ol}, and \code{mlc2ol}, 
which use the data points on the left or right side of 
the peak accordingly in ``in'' and ``out'' time domain.
We apply the selection criteria that all of these chi-squared values 
are smaller than $3.0$.

A real microlensing light curve should be symmetric with respect to 
its peak time \footnote{Note this is not the case in general 
when we consider the binary lens, parallax, etc.}. 
We quantify the degree of asymmetry in the measured light curve by
\begin{align}
    \code{asym} = \frac{\sum_{i\in \text{pairs}} w_i |\Delta F_{i+}-\Delta F_{i-} |}{\sum_{i\in \text{pairs}} w_i (\Delta F_{i+} + \Delta F_{i-} - 2\min\Delta F)}.
    \label{eq:asym}
\end{align}
Here, each pair of time indices ($i+,i-$) is chosen symmetrically 
around the peak time $t_0$; 
the summation includes only the pairs that satisfy $|t_{i\pm}-t_0|\leq t_{\rm FWHM}$. 
We apply a selection criterion, requiring $\code{asym}\leq0.17$, 
to exclude light curves with significant asymmetry, as such events are 
unlikely to be due to microlensing.

We next consider a selection to remove light curves affected by 
weather conditions. To quantify the impact of weather conditions,
we compute the Pearson's correlation coefficient between 
the light curve and the seeing, 
using the latter as a proxy of the weather conditions:
\begin{align}
    \code{scorr} = \frac{\sum_i w_i(\Delta F_i - \overline{\Delta F}) (S_i - \overline{S})}{ \sqrt{\sum_i w_i (\Delta F_i-\overline{\Delta F})^2} \sqrt{\sum_i w_i (S_i-\overline{S})^2}}
    \label{eq:scorr}
\end{align}
where $S_i$ is the seeing at the $i$-th image and we use 
the joint inverse variance weight $w_i = 1/[\sigma(\Delta F)^2\times \sigma(s)^2]$.
The sum runs over the images within FWHM time interval centered on the peak. 
We apply a selection criterion that requires 
$\code{scorr}\leq 3.0/\sqrt{\rm dof}$, where ${\rm dof}$ is the number of 
degrees of freedom used to evaluate \code{scorr}.

The above definition of the correlation captures the correlation between the
light curve and the seeing over all time scales.
However, a light curve may also exhibit correlations on shorter timescales. 
To account for such correlations, we introduce a timescale dependent 
seeing correlation, which uses the same definition as Eq.~(\ref{eq:scorr}) 
but is computed using only the data points within a specific time 
interval corresponding to a fixed time range $t_{i_{\rm max}} - t_{i_{\rm min}}$. 
Then we average over all possible choices of $i_{\rm max}$ to define 
the seeing correlation for a fixed timescale, \code{scorrs}, 
together with its associated mean degree-of-freedom. 
Our selection criteria is that all time-dependent seeing correlations must be smaller 
than $3\sigma$; that is, $\max\left[\code{scorrs}/\sqrt{\rm dof}\right]\leq3.0$.

For secure event selection, we also require the microlensing event to be significant. 
The first selection criterion is that the maximum magnification, $\max{A}$, 
predicted by the best-fit microlensing model parameters must exceed $1.34$. 
We call this selection as \code{amax}.
However, this criterion relies on the parameters derived from the best-fit model 
and does not account for the uncertainties in the light curve. 
To incorporate the uncertainties, we adopt the following criteria for the significance:
\begin{align}
    \code{mlsig} = \sum_i \left(\frac{\Delta F_i - \min[\Delta F_i^{\rm model}]}{\sigma(\Delta F_i)}\right)^2,
    \label{eq:mlsig}
\end{align}
where $\Delta F_i$ is the light curve data at the $i$-th epoch, 
${\rm min}[\Delta F_i^{\rm model}]$ is the minimum flux for 
the best-fit microlensing model over the interval of observational time, 
and $\sigma(\Delta F_i)$ denotes the observational error at the $i$-th epoch. 
We select the events requiring $\code{mlsig}/{\rm dof}>5$.
This criterion can be interpreted that the data points have
$\sqrt{5}\sigma=2.23\sigma$ deviation from no-variation on average.
The reason we use the minimum flux from the best-fit model is that 
the observed light curve is sometimes too noisy to robustly estimate the minimum flux.

Table~\ref{tab:selection-result} summarizes the results of event selection 
at each selection stage. 
After all of these first-level selection steps, we retained 38, 18, and 8 events 
from 2014-11-24, 2017-09-20, and 2020-11-24 datasets, respectively. 
For these remaining events, we performed a visual 
inspection as the second-level selection. 
In this selection, we looked into the raw and difference images around 
the detected position, to assess whether the event is consistent with microlensing 
or instead an impostor, e.g. due to a moving object or imperfect image subtraction.
We denote this visual selection as ``mlvs'' in the Table~\ref{tab:selection-criteria}. 
We identified several impostors, and were left with 25, 18, and 7 candidate events 
in the 2014-11-24, 2017-09-20, and 2020-11-24 datasets, respectively.

The microlensing event rate per star is very small, so it is highly unlikely that 
the same star would undergo multiple microlensing events.
Therefore, repeated variations for the same star (or, more precisely, at the same position) 
observed on different dates are very likely due to an intrinsic variable star. 
To ensure that the events are not due to repeating variables, 
we performed the forced photometry at the position of selected events 
on different observation dates. We examined both the light curves and the images, 
and removed the events showing suspicious variations in the light curve 
across different dates. 
However, note that we did not remove the events if the apparent variation 
on a different observation date is attributable to poor image subtraction. 
This selection is labeled as ``mlnrpt'' in Table~\ref{tab:selection-criteria}. 
After this final selection, we were left with 5 and 7 candidate events 
in 2014-11-24, and 2017-09-20 datasets, respectively. 
These 12 events constitute the sample of microlensing candidates 
used in the following results.

\subsection{Candidate and secure catalogs}
\label{ssec:candidate-secure-catalog}
\begin{figure}
    \includegraphics[width=0.45\textwidth]{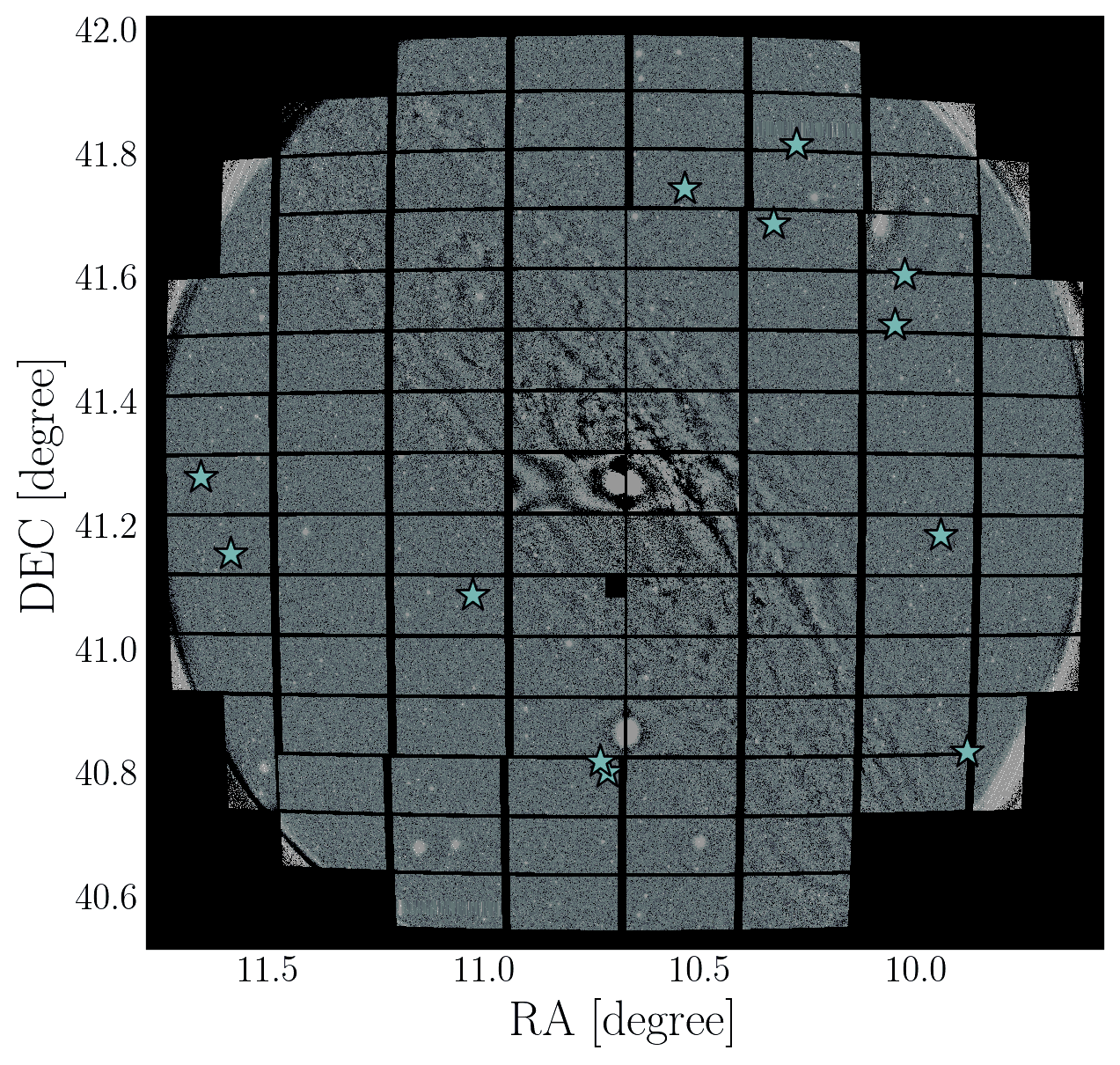} 
    \caption{
    An example of the image  for the M31 region, 
    where the reference image obtained under 
    the best seeing conditions during a given night is subtracted 
    from a target image.
    Most stars in the spiral disk regions of M31 are well subtracted. 
    Point sources remaining in the difference image 
    are candidates of variable stars. Cyan star symbols denote the locations 
    of the 12 microlensing candidates that passed all the selection criteria.
    }
    \label{fig:spacial-event-dist}
\end{figure}

Table~\ref{tab:candidate-catalog} summarizes the properties of 
the selected events, which we refer to the {\it candidate} catalog.
Appendix~\ref{apdx:light-curve} also shows 
the light curves of these events. 
For each candidate event, we run an additional parameter 
inference using \code{MultiNest} 
\cite{Feroz.Bridges.2009} to obtain an accurate posterior distribution.
In Appendix~\ref{apdx:lc-post},
we show the posterior distribution of the 
PLFS model parameters for each candidate. 
Note that the degeneracies among the model parameters are 
generally strong, and the parameter values shown 
in Table~\ref{tab:candidate-catalog} 
are just a nominal set of parameters.

We notice that the light curves of some candidates have peaks at 
at similar times during the observing night.
In the simulation, we checked that the events around these times are
more likely to be detected. 
This can be understood as follows. 
Since the seeing tends to be worse near the start or the end 
of the observation on each night, as shown in Fig.~\ref{fig:seeing}, 
a light curve whose peak occurs at those times exhibits an 
apparently higher correlation with seeing.
Such events 
are therefore removed by the selection criteria based on 
the correlations with seeing.

In Fig.~\ref{fig:spacial-event-dist}, we show the locations
of the 12 microlensing candidates overlaid on 
an HSC image of the M31 region.

We also define the {\it secure} catalog by further requiring 
$\code{mlsig}/{\rm dof}>10$, for which we intend to select 
the event at higher significance although the criterion is heuristic. 
The events in the secure catalog are marked by $\dagger$ 
in Table~\ref{tab:candidate-catalog}.

\section{Efficiency characterization}\label{sec:efficiency}
\begin{table}
    \setlength{\tabcolsep}{8pt}
    \centering
    \caption{
    Ranges of parameters used to simulate microlensing light curves for 
    estimating the detection efficiency. 
    All parameters are randomly sampled from the listed intervals 
    to simulate each light curve.
    }
    \begin{tabular}{cl}
    \hline\hline
    Parameter & Prior \\
    \hline
    $u_0$ & flat$[0.0, 2.0]$ \\
    $t_0$ & flat$[0.0,t_{\rm obs}]$ \\
    $\log_{10}(t_{\rm E}/t_{\rm obs})$ & flat$[\log_{10}(2\,{\rm min}
    /t_{\rm obs}/10.0), \log_{10}(2.0)]$ \\
    $\rho$ & flat$[0.0, 2.0]$\\
    \hline\hline
    \end{tabular}
    \label{tab:simulation-prior}
\end{table}

\begin{table}
    \setlength{\tabcolsep}{8pt}
    \centering
    \caption{Number of source stars, $N_{{\rm s}, m}$, per magnitude bin 
    $m$ estimated in \citet{Niikura.Chiba.2019}.}
    \begin{tabular}{cr}
    \hline\hline
    Magnitude & $N_{{\rm s}, m}$ \\
    \hline
    22 & 675,465 \\
    23 & 2,326,642 \\
    24 & 11,827,524 \\
    25 & 36,627,840 \\
    26 & 31,793,192 \\
    \hline\hline
    \end{tabular}
    \label{tab:source-number}
\end{table}

\begin{figure*}
    \includegraphics[width=\textwidth]{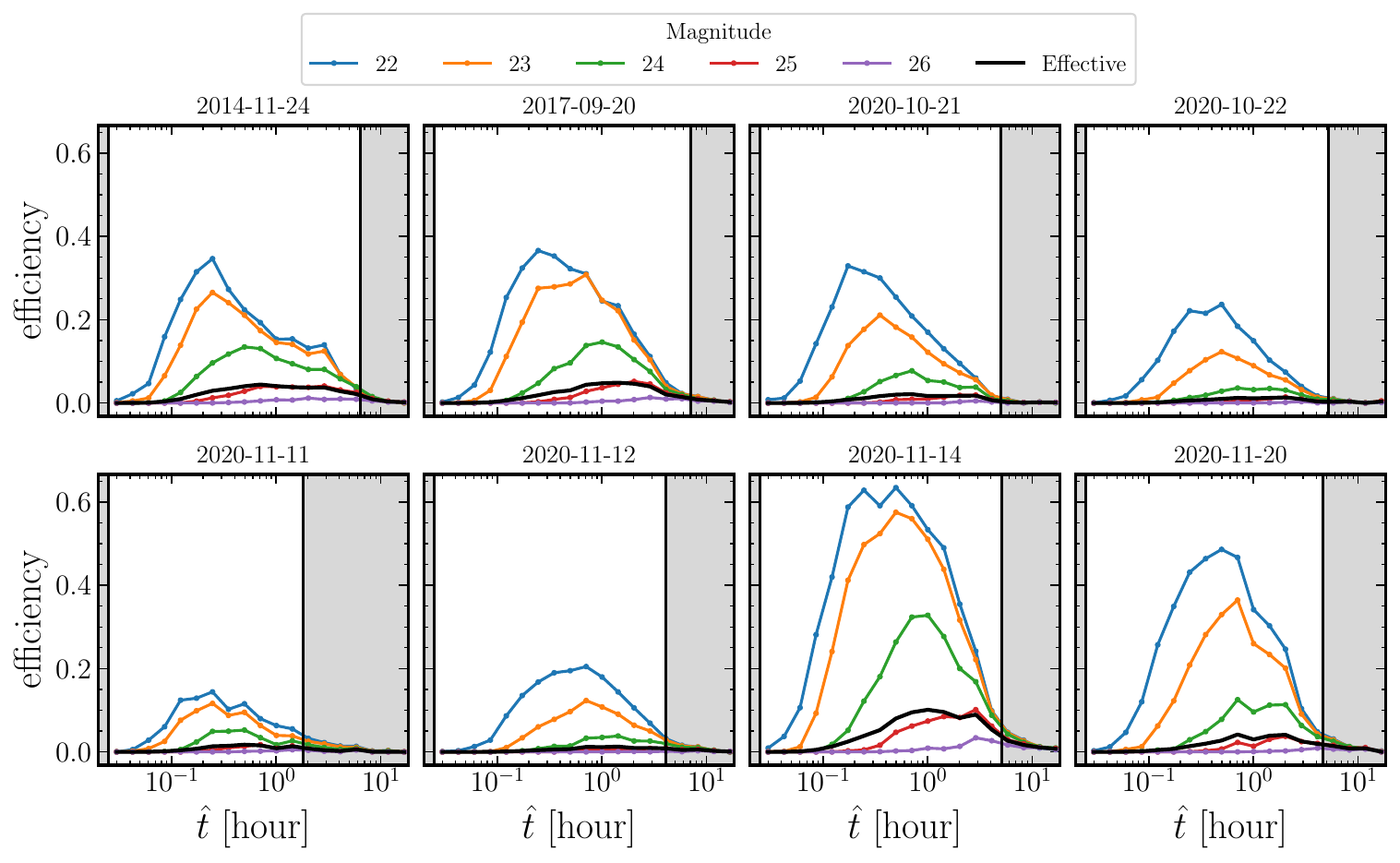} 
    \caption{
    The detection efficiency of the microlensing events as a function of the 
    microlensing light-curve timescale $\hat{t}\equiv t_{\rm E}\sqrt{u_T^2(\rho)-u_0^2}$.
    Each panel shows the detection efficiency for 
    each observation data, as indicated in the title of panel. 
    Different colors show the efficiency for source stars with different 
    intrinsic brightness, i.e., $(22,23,24,25,26)$~mag, respectively. 
    For each combination of observation date, source brightness, and HSC patch, 
    we simulated 10,000 realizations of microlensing light curves by
    randomly drawing model parameters from the priors listed
    in Table~\ref{tab:simulation-prior} and adding 
    Gaussian random noise whose amplitude is 
    estimated from the corresponding
    data in each HSC patch and on each observation date. 
    The black curve in each panel indicates the {\it effective} efficiency, 
    obtained by integrating the brightness-dependent efficiencies weighted by 
    the brightness distribution of source stars in M31, as
    defined in Eq.~(\ref{eq:effective-efficiency}).
    The shaded regions in each panel indicate the regions of 
    microlensing timescales that are difficult to detect in each dataset: 
    for each observation data, 
    the upper timescale cut corresponds to $T_{\rm eff}$ listed 
    in Table~\ref{tab:obs-summary}.
    }
    \label{fig:efficiency}
\end{figure*}

To compare the observed number of microlensing events with theoretical expectations, 
it is necessary to characterize the detection efficiency of microlensing events 
in the observed data. 
The efficiency is defined as the fraction of underlying microlensing events 
that are detected by our analysis pipeline. 
The efficiency depends on both the properties of individual microlensing events 
and the observational conditions.

We use simulations to estimate the detection efficiency. 
In this paper, we simulate microlensing events at the light-curve level 
rather than at the image level. 
This method was validated by \citet{Niikura.Chiba.2019}, which demonstrated 
that the efficiencies estimated from light-curve-level simulations are 
consistent with those obtained from image-level simulations.
For every combination of observation date/time and HSC patch, 
we simulate 10,000 light curves for PLFS microlensing events:
\begin{align}
    \Delta F(t_i) &= F(t_i) - \frac{1}{10} \sum_{j\in{\rm ref}}F(t_j) + N_i ,
    \label{eq:df-sim}
\end{align}
with 
\begin{align}
    F(t_i) &= F_0 A(t_i;\bm\theta) .
\end{align}
Here $\bm{\theta} = (u_{0}, t_{0}, t_{\rm E}, \rho)$ is the set of 
microlensing model parameters, each drawn uniformly from the prior range 
in Table~\ref{tab:simulation-prior}. 
For the intrinsic flux $F_0$, we adopt discrete values in photon counts:
$\{10^{0.4\times5}, 10^{0.4\times4}, 10^{0.4\times3}, 10^{0.4\times2}, 10^{0.4\times1}\}$ 
corresponding to $\{22,23,24,25,26\}$~mag, respectively.
The second term $N_i$ in Eq.~(\ref{eq:df-sim}) is the noise term, which is 
drawn from a normal distribution with zero mean and a standard deviation 
estimated by averaging the errors of all light curves in 
the corresponding master catalog in the patch under consideration. 
After running simulations in each magnitude bin $m$, 
we estimate the efficiency $\epsilon_m^n(\hat{t})$ as 
the ratio of the number of detected events to the number of injected 
events for each microlensing timescale 
$\hat{t}\equiv t_{\rm E}\sqrt{u_T^2(\rho) - u_0^2}$ for each night $n$.
After running the entire process described above, we found
that the efficiency is low and noisy in some magnitude bins; 
therefore, we performed an additional set of simulations with 
four times as many realizations.

We also define the effective efficiency by averaging the efficiencies in 
different magnitude bins, weighted by the number of background source stars:
\begin{align}
    \epsilon_{\rm eff}^n(\hat{t}) = \frac{1}{N_{\rm s, tot}}\sum_{m} 
    \epsilon_m^n(\hat{t}) N_{{\rm s}, m},
    \label{eq:effective-efficiency}
\end{align}
where $N_{{\rm s}, m}$ is the number of source stars in the magnitude bin $m$, 
and $N_{\rm s, tot}$ is the sum of them. We use the 
numbers of source stars in each magnitude bin, 
estimated in Ref.~\citep{Niikura.Chiba.2019}, as
summarized in Table~\ref{tab:source-number}.

Fig.~\ref{fig:efficiency} shows the estimated detection efficiency as a function 
of the microlensing light curve timescale, $\hat{t}$. 
We first note that the efficiency clearly depends on the observation date, 
reflecting varying observing conditions. In particular, the efficiencies for 
2014-11-24, 2017-09-20, and 2020-11-14 are higher than the other dates, 
consistent with expectations from the typical seeing on those nights, 
as summarized in Table~\ref{tab:obs-summary}. 
However, we suspect that the efficiency for 2020-11-14 night would be overestimated,
because we did not detect any microlensing candidates on that night. 
We think that this is due to the relatively poor seeing conditions compared to 
those on the nights of 2014-11-24 and 2017-09-20, during which 
the microlensing candidates were detected, as can be found from Fig.~\ref{fig:seeing}.
In Appendix~\ref{apdx:bad-seeing-reference}, 
we checked that the efficiency of microlensing candidate detection is sensitive to 
the seeing conditions of the reference image used in the image difference technique.
Therefore, the relatively poor seeing on 2020-11-14 limits the sensitivity of 
microlensing detection. However, the light curve simulations do not adequately 
reproduce the degraded uncertainties. This is beyond the scope of this paper, 
and we will revisit it in the future. 
In this paper, we include the microlensing detection efficiency for
2020-11-14 in the following results.

We also note a non-zero efficiency even for light curve timescales longer than 
the observational duration of each night ($\sim$7~hours).
This is because partial light-curve coverage can still be fit within the 
statistical errors by different combinations of the parameters, owing to 
parameter degeneracies. 
Consequently, some of such simulated light curves can pass 
the selection criteria and be detected.

\section{Likelihood modeling for population parameter inference}
\label{sec:likelihood-model}
\begin{figure*}
\includegraphics[width=\textwidth]{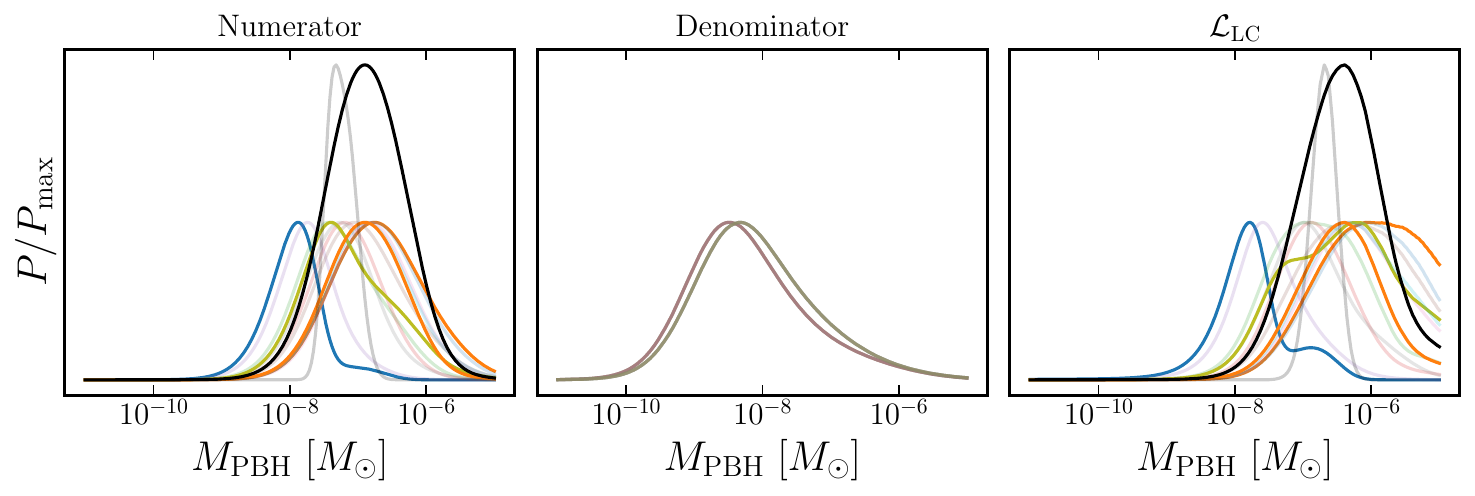} 
    \caption{The rightmost panel shows the profiles of the light-curve likelihoods 
    for each of the 12 microlensing candidates, computed using Eq.~(\ref{eq:lc-like}). 
    Here we adopt the monochromatic PBH mass function defined in 
    Eq.~(\ref{eq:mono-mass-func}). 
    The colored lines are the likelihoods for individual candidates
    $\mathcal{L}_{\rm LC}(\bm{d}_j^n|M_{\rm PBH})$, 
    each normalized by their maximum values.
    The bold lines show the likelihoods for 
    the 4 secure candidates (see Table~\ref{tab:candidate-catalog}).
    The joint likelihood $\mathcal{L}_{\rm LC}(\{\bm{d}_j^n\}|M_{\rm PBH})$ 
    for all the candidates or for the 4 secure candidates
    is shown by the thin gray or bold black line, respectively, and each profile is
    normalized such that its peak is twice as high as those of the individual likelihoods.
    The left and middle panels show the numerator and denominator of 
    the light-curve likelihoods. 
    Note that, since the candidates are obtained
    only on 2014-11-24 and 2017-09-20, the middle panel contains only two lines.
    }
    \label{fig:lclike-profile}
\end{figure*}

As described in Section~\ref{sec:data-analysis}, we retain 5 and 7 
microlensing candidate events from the 2014-11-24 and 2017-09-20 data, 
while no events are selected from the other observing dates. 
Additionally, we obtained the posterior distribution of 
the microlensing parameters for each of these candidate microlensing events. 
In this section, we describe how we constrain the parameters of 
the PBH population model, $\bm{\lambda}$,  from these observed data.

The simplest way is to construct the Poisson likelihood for 
the observed number of the events and compare it with 
the theoretical prediction based on the mass function under consideration:
\begin{align}
    \mathcal{L}_{\rm Po}\left(\{N_{{\rm obs}, n}\} | \bm{\lambda}\right) = 
    \prod_{n=1}^{8} P_{\rm Po}(N_{{\rm obs}, n} | N_{{\rm exp}, n}(\bm{\lambda}))
    \label{eq:poisson-like}
\end{align}
where $P_{\rm Po}(N_{{\rm obs}}| N_{{\rm exp}})$ is the Poisson probability 
of observing $N_{\rm obs}$ events given an expectation of $N_{\rm exp}$ events, 
and $N_{{\rm obs}, n}$ and $N_{{\rm exp}, n}$ are the observed and 
expected numbers of events for each night indexed by $n$. 
The expected number of events depends on the PBH population parameters $\bm{\lambda}$. 
In the next section, we will consider several working hypotheses to 
infer the PBH population parameters.
Since we consider different values of $N_{\rm obs}$, depending on 
the level of confidence in the candidates,
we do not specify the value of $N_{{\rm obs}, n}$ in this section.
Because the Poisson probability only tests the number of 
the observed events, it is only sensitive to the abundance parameter, 
and cannot be used to infer other population parameters such as mass scale.

To constrain PBH population parameters beyond the abundance parameter,
we employ hierarchical Bayesian inference that incorporates 
the light curve information for each microlensing 
candidate, in addition to the number of microlensing events for each night.
If the microlensing light curve has a sufficient signal-to-noise ratio,  
the PLFS parameters can be estimated precisely, which in turn can be used 
to infer the lens mass, i.e., the PBH mass~\cite{Sugiyama.Kusenko.2021}.
However, the different candidates have different precisions in 
the estimation of the PLFS parameters,
as shown in Appendix~\ref{apdx:lc-post}.
We employ the method in Ref.~\cite{Mandel.Gair.2019} to combine 
the light curve information from the different candidates 
\citep[see also][for earlier work on this problem]{Loredo.Loredo.2004}.

For notational simplicity, we denote the $j$-th light curve data in 
the $n$-th night by the vector $\bm{d}_j^n=\{\Delta F_i\}$ in the following.
The total likelihood for a set of the light curve data $\{\bm{d}_j^n\}$ can 
be modeled, following Eq.~(7) of Ref.~\cite{Mandel.Gair.2019}, as
\begin{align}
    \mathcal{L}_{\rm LC}(\{\bm{d}_j^n\} | \bm{\lambda}) &= 
    \prod_{n=1}^8\prod_{j=1}^{N_{{\rm obs},n}} 
    \mathcal{L}_{\rm LC}(\bm{d}_j^n | \bm{\lambda}) 
\end{align}
with
\begin{align}
    \mathcal{L}_{\rm LC}(\bm{d}_j^n | \bm{\lambda}) &= 
    \frac{
    \int\dd d
    \int\dd\bm{\theta} P(\bm{d}_j^n | \bm{\theta}) 
    P_{\rm pop}(\bm{\theta},d | \bm{\lambda})
    }{
    \int\dd d
    \int\dd\bm{\theta}\epsilon_{\rm eff}^n(\bm{\theta}) 
    P_{\rm pop}(\bm{\theta},d|\bm{\lambda})
    },
    \label{eq:lc-like-original}
\end{align}
where $P(\bm{d}_j^n | \bm{\theta})$ is the likelihood of observing
the light curve $\bm{d}_j^n$ for a given set of PLFS parameters 
$\bm{\theta}$ on the $n$-th night, and $\epsilon_{\rm eff}^n(\bm{\theta})$ 
is the detection efficiency for that night,
given as a function of $\bm{\theta}$. 
We note that the population model, specified by $\bm{\lambda}$,  
also provides the probability distribution for PBHs at the distance $d$, 
but the light curve itself is not sensitive to $d$
because of parameter degeneracies.
Therefore, we assume the likelihood of observing the light curve depends 
only on $\bm{\theta}$, not on $d$.
We also note that the PLFS parameters $t_0$ and $f_0$ do not contain
any information about the population, and thus we implicitly marginalize 
over these variables and assume $\bm{\theta}=\{t_{\rm E}, u_0, \rho\}$.
The population distribution $P_{\rm pop}(\bm{\theta}|\bm{\lambda})$ describes 
the distribution of events for a given set of the population parameters 
$\bm{\lambda}$, and is defined by the normalized differential event rate:
\begin{align}
    \frac{\dd\Gamma}{\dd d\dd\bm{\theta}}(\bm{\lambda}) 
    = N(\bm{\lambda}) P_{\rm pop}(\bm{\theta} ,d|\bm{\lambda})
\end{align}
with $\int\dd\bm{\theta}P_{\rm pop}(\bm{\theta}|\bm{\lambda})=1$.
See Appendix~\ref{apdx:eventrate-model} for the detailed definition of 
the differential event rate.

By multiplying $N(\bm{\lambda})$ to both the numerator and denominator of 
Eq.~(\ref{eq:lc-like-original}), we can replace the population probability 
by the differential event rate:
\begin{align}
    \mathcal{L}_{\rm LC}(\bm{d}_j^n | \bm{\lambda}) = 
    \frac{
    \int\dd d
    \int\dd\bm{\theta} P(\bm{d}_j^n | \bm{\theta}) 
    \frac{\dd\Gamma}{\dd d\dd\bm\theta}(\bm{\lambda})
    }{
    \int\dd d
    \int\dd\bm{\theta}\epsilon_{\rm eff}^n(\bm{\theta}) 
    \frac{\dd\Gamma}{\dd d\dd\bm\theta}(\bm{\lambda})
    } .
    \label{eq:lc-like}
\end{align}
Then the integral in the denominator yields the standard event rate as 
a function of the population parameters.
In the previous section, we checked that the efficiency mainly depends 
on the PLFS parameters through $\hat{t}$, and therefore the denominator 
of Eq.~(\ref{eq:lc-like}) can be simplified as
\begin{align}
    \Gamma_n(\bm{\lambda}) = \
    \int\dd d\int\dd\hat{t}~\epsilon_{\rm eff}^n(\hat{t})
    \frac{\dd\Gamma}{\dd d\dd\hat{t}}(\bm{\lambda}).
    \label{eq:int-ep-ppop}
\end{align}
The detailed form is given by Eq.~(\ref{eq:eventrate}).

In Section~\ref{ssec:candidate-secure-catalog} and Appendix~\ref{apdx:lc-post}, 
we show that the PLFS parameters $\bm{\theta}$ have strong degeneracies 
among themselves in their posterior distributions. 
This means that the multidimensional integral in the numerator of 
Eq.~(\ref{eq:lc-like}) is difficult to reduce to a lower-dimensional integral. 
By following \citet{Mandel.Gair.2019} and using appropriate changes of variables, 
we evaluate the integral in the numerator using the Monte-Carlo integration 
with the posterior samples obtained from the light curve fitting.
We leave the details of the computation to Appendix~\ref{apdx:lc-like-mc-integral}.

To simultaneously constrain the abundance and other population parameters of PBHs 
(e.g., their mass and mass function), we combine the Poisson likelihood in 
Eq.~(\ref{eq:poisson-like}) with the light-curve likelihood in Eq.~(\ref{eq:lc-like}):
\begin{align}
    \mathcal{L}(\{\bm{d}_j^n\}, \{N_{{\rm obs},n}\} | \bm{\lambda}) =
    \mathcal{L}_{\rm Po}(\{N_{{\rm obs},n}\} | \bm{\lambda}) 
    \times \mathcal{L}_{\rm LC}(\{\bm{d}_j^n\} | \bm{\lambda}).
    \label{eq:joint-like}
\end{align}
From Bayes theorem, we obtain the posterior distribution of $\bm{\lambda}$ 
for given datasets
\begin{align}
    \mathcal{P}_{\rm Po}(\bm{\lambda}| \{N_{{\rm obs},n}\})
    &= \mathcal{L}_{\rm Po}(\{N_{{\rm obs},n}\} | \bm{\lambda}) \Pi(\bm{\lambda})
    \label{eq:poisson-posterior}
\end{align}
or
\begin{align}
    \mathcal{P}(\bm{\lambda}| \{\bm{d}_j^n\}, \{N_{{\rm obs},n}\})
    &= \mathcal{L}(\{\bm{d}_j^n\}, \{N_{{\rm obs},n}\} | \bm{\lambda}) 
    \Pi(\bm{\lambda})
    \label{eq:joint-posterior}
\end{align}
up to a normalization constant, respectively.
The former (Eq.~(\ref{eq:poisson-posterior})) corresponds to the result obtained 
using only the number of microlensing events, while the latter 
(Eq.~(\ref{eq:joint-posterior})) corresponds to the result obtained by 
combining the number of events with the light-curve information for each candidate.

Fig.~\ref{fig:lclike-profile} shows the profiles of the light-curve likelihoods for 
the individual events, as well as their joint profile, 
assuming a monochromatic mass function for PBHs:
\begin{align}
    \frac{{\rm d}n}{{\rm d}\ln M}=
    f_{\rm PBH}\delta^{\rm D}(\ln M-\ln M_{\rm PBH})
    \label{eq:mono-mass-func},
\end{align}
where the population parameters are $\bm{\lambda}=\{f_{\rm PBH}, M_{\rm PBH}\}$.
For the monochromatic mass function, the light-curve likelihoods do not depend 
on the abundance parameter $f_{\rm PBH}$, and only depends on $M_{\rm PBH}$; 
$\mathcal{L}_{\rm LC}(\{\bm{d}_j^n\}|\bm{\lambda})=\mathcal{L}_{\rm LC}(\{\bm{d}_j^n\}|M_{\rm PBH})$. 
Therefore, the profiles of the light-curve likelihoods is shown as functions of 
$M_{\rm PBH}$ in the figure.

The light-curve likelihood for a more generic mass function 
$\frac{\dd n}{\dd\ln M}(M;\bm{\lambda})$ can be easily obtained by 
extending the above method.
We convolve the numerator and denominator of the light-curve likelihood 
separately with the mass function, and then take the ratio of the 
convolved quantities to obtain the desired likelihood.

\section{Results}
\label{sec:result}

In this section, we constrain the abundance of compact objects, such as PBHs, 
in the dark matter halos of the Milky Way and M31, using the likelihoods 
defined in the previous section.
We consider three different working hypotheses.

\subsection{Null hypothesis: Upper limit on the PBH abundance}\label{sec:upperlim}
\begin{figure}[t]
    \includegraphics[width=0.45\textwidth]{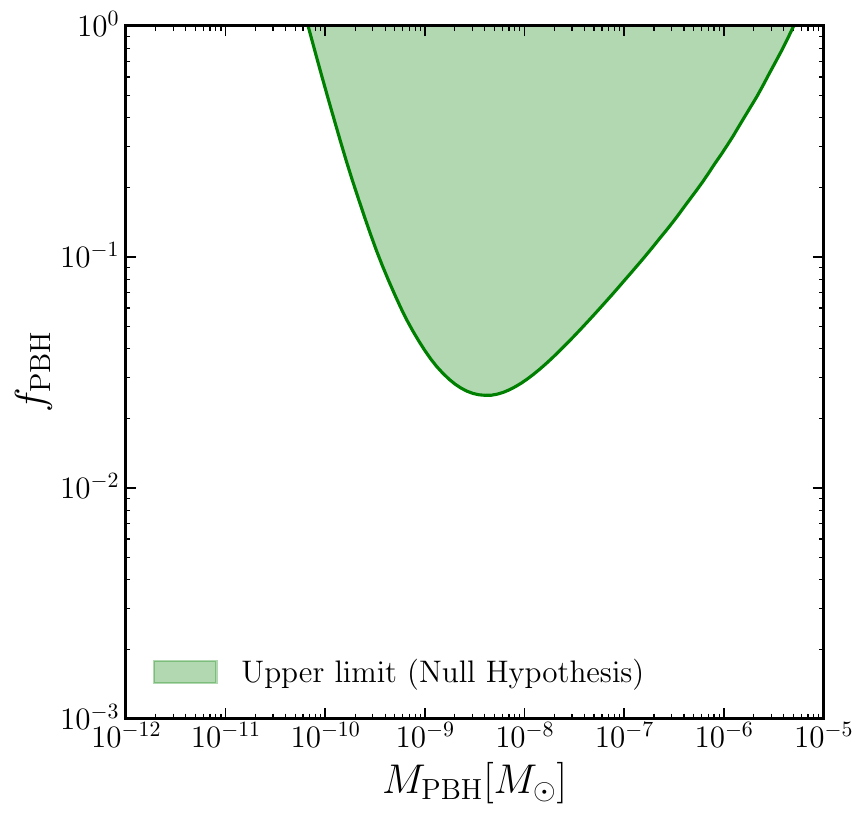}
    \caption{
    The 95\% C.L. exclusion region of
    the PBH abundance parameter $f_{\rm PBH}$ for each 
    mass scale of $M_{\rm PBH}$ (green region).
    Here, $f_{\rm PBH}$ is the PBH mass
    fraction to the total dark matter in the 
    Milky Way and M31  halo regions.
    We obtained the region assuming 
    a monochromatic mass function under 
    the null hypothesis, 
    namely that all our candidates are false positives.}
    \label{fig:upperlim}
\end{figure}

In the null hypothesis, we assume that all the candidate events are 
false positive caused by, e.g. variable stars, rather than by microlensing 
due to PBH.
Under this hypothesis, we derive an upper limit on the abundance of compact objects, 
such as PBHs, as a component of dark matter.
More specifically, we use the Poisson likelihood defined in 
Eq.~(\ref{eq:poisson-like}) by setting
$N_{{\rm obs},n}=(0,0,0,0,0,0,0,0)$.
In this hypothesis, we consider the monochromatic PBH mass function, 
defined in Eq.~(\ref{eq:mono-mass-func}), 
and derive an upper bound on $f_{\rm PBH}$ as a function of PBH mass scale 
$M_{\rm PBH}$ by imposing ${\cal L}>0.05$, 
which corresponds to 95\% confidence level (C.L.).
This means that we treat $\bm{\lambda}=\{f_{\rm PBH}\}$ as a model parameter 
and constrain it at each of the varied mass scales $M_{\rm PBH}$.

Figure~\ref{fig:upperlim} shows the upper limit.
The constraint is not improved compared to \citet{Niikura.Chiba.2019} or
the subsequent work~\citet{Smith.Consortium.2002}, 
even though we use a larger data under the same hypothesis. 
This is because we include the finite source size effect 
in our analysis pipeline, 
which lowers the efficiency, since the microlensing events 
with finite source-star size generally 
produce smaller magnifications.

\subsection{PBH hypothesis: Posteriors for the PBH mass function parameters}
\begin{figure*}
    \includegraphics[width=0.48\textwidth]{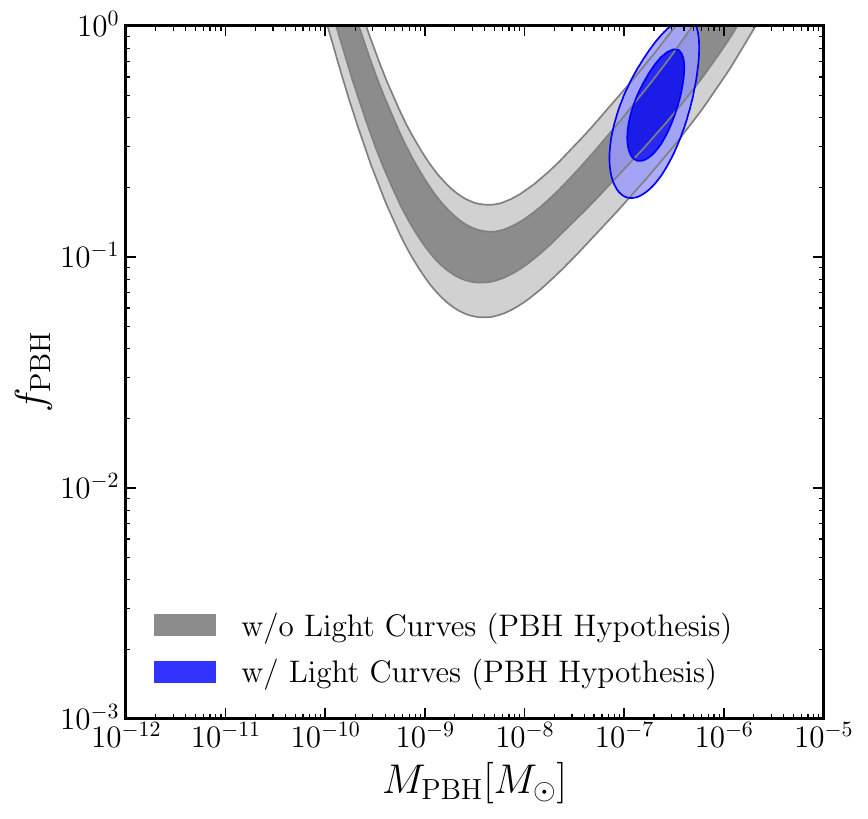}
    \includegraphics[width=0.48\textwidth]{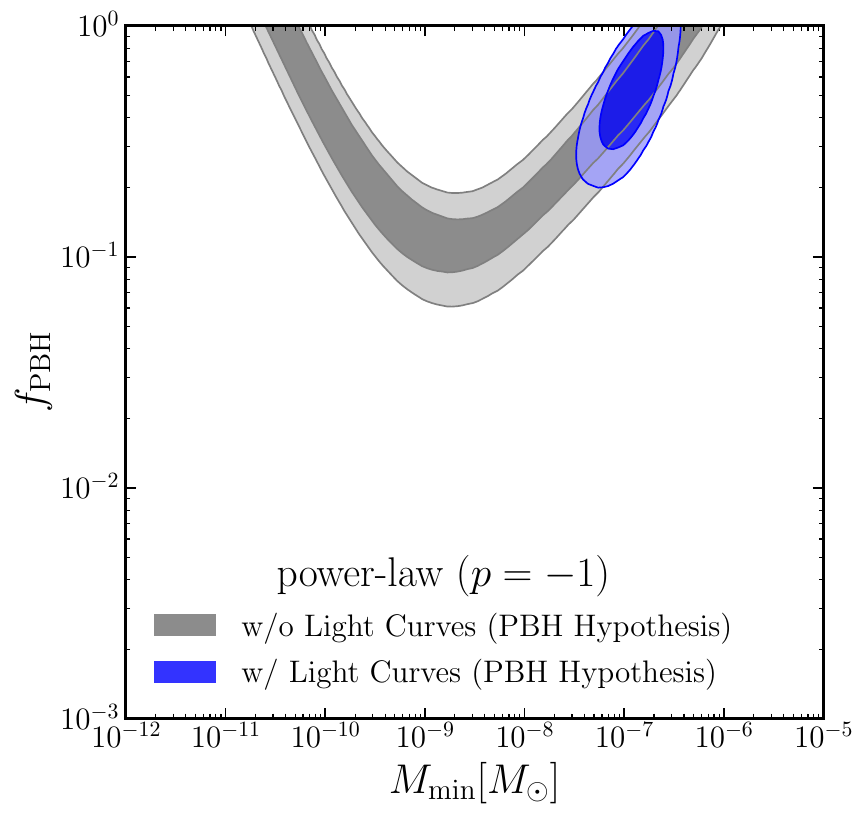}
    \caption{The shaded regions show the 68\% and 95\% C.L. allowed regions of the 
    PBH abundance parameters
    under the PBH hypothesis, 
    where we assumed that all the candidates 
    are due the to PBH microlensing.
    \textit{Left}: The posterior distributions obtained assuming
    the monochromatic PBH mass function.
    The gray contours are obtained using
    the Poisson likelihood for the number of candidates. In this case 
    we obtain the allowed range only for $f_{\rm PBH}$ at each PBH mass scale.
    The blue contours are obtained using the likelihood that further incorporates
    the light-curve information for each candidate (Fig.~\ref{fig:lclike-profile}) 
    together with the Poisson likelihood. 
    In this case, the PBH mass scale is also constrained simultaneously with $f_{\rm PBH}$.
    \textit{Right}: Similar results, but assuming
    the power-law PBH mass function defined in Eq.~(\ref{eq:pl-massfunc}).
    We considered the fixed power-law index $p=-1$ for the mass function 
    $\mathrm{d}n_{\rm PBH}/\mathrm{d}\ln M_{\rm PBH}\propto M^{p}$, 
    where $M_{\rm min}$ is a parameter that characterizes the minimum PBH mass scale.} 
    \label{fig:plausible}
\end{figure*}

In this section, we consider PBH hypothesis, where we assume that 
all the events in the {\it candidate} catalog are caused by PBHs, 
with probabilities given by the likelihood function. 
This hypothesis allows us to derive posterior distributions 
for the PBH abundance parameters, or more generally for parameters 
that characterize PBH mass function, 
using the candidate catalog summarized in Table~\ref{tab:candidate-catalog}. 
In this case, we set $N_{{\rm obs},n}=(5,7,0,0,0,0,0,0)$.

We start with deriving the allowed region of 
$\bm{\lambda}=(f_{\rm PBH}, M_{\rm PBH})$ 
for the monochromatic mass function (Eq.~(\ref{eq:mono-mass-func})),
based on the PBH hypothesis. 
We first derive the parameter constraints using the likelihood 
for the number of microlensing events, as
defined in Eq.~(\ref{eq:poisson-posterior}).
We adopt uniform priors on $\ln f_{\rm PBH}$ and $\ln M_{\rm PBH}$.
The gray contours in Fig.~\ref{fig:plausible} show
the resulting constraints on ($f_{\rm PBH}, M_{\rm PBH}$).
Because we use only the number of detected events, the mass scale 
is not constrained at all, 
and only the abundance parameter $f_{\rm PBH}$ is constrained 
at each mass scale $M_{\rm PBH}$. 
This means that the parameters in the gray bands predict 
nearly identical event rates and therefore cannot be distinguished 
only through the observed number of events.

Next, we examine the case in which the light-curve information 
for each candidate is incorporated together with the number of events, 
as defined in Eq.~(\ref{eq:joint-posterior}). 
The blue contours in Fig.~\ref{fig:plausible} show the allowed region
of the PBH mass function parameters, $(f_{\rm PBH}, M_{\rm PBH})$. 
It is clear that the constraint on mass scale is 
significantly improved, compared to the gray contours.

Motivated by the constraining power on the PBH mass scale 
demonstrated above, we also consider the case in which PBHs 
follow the power-law mass function defined by
\begin{align}
    \frac{\dd n}{\dd\ln M} = 
    f_{\rm PBH} p\left(\frac{M}{M_{\rm min}}\right)^p\Theta(M- M_{\rm min}),
    \label{eq:pl-massfunc}
\end{align}
where $p<0$ is the power-law index, for which we fix $p=-1$, 
$\Theta(x)$ is the Heaviside function, defined
as $\Theta(x)=1$ if $x>0$ and $\Theta(x)=0$ otherwise, 
and $M_{\rm min}$ is the minimum PBH mass scale. 
This model is normalized such that the total abundance becomes 
$f_{\rm PBH}$ after integrating over $M_{\rm PBH}$.
The model parameters are $\bm{\lambda}=(f_{\rm PBH}, M_{\rm min})$. 
We adopt uniform priors on $\ln f_{\rm PBH}$ and $\ln M_{\rm min}$.

The right panel of Fig.~\ref{fig:plausible} shows 
the allowed region for these parameters. 
We notice that the contours look similar to those for the
monochromatic mass function, but are slightly shifted toward lower masses. 
This is because the power-law mass function produces microlensing events 
from PBHs at mass scales greater than
the minimum mass scale, $M>M_{\rm min}$.

In Appendix~\ref{apdx:test-model}, we present some tests of more complicated and extended mass functions, which 
are motivated by the theory of PBH formation.

\subsection{PBH hypothesis with the secure microlensing candidates}
\begin{figure*}
    \includegraphics[width=0.48\textwidth]{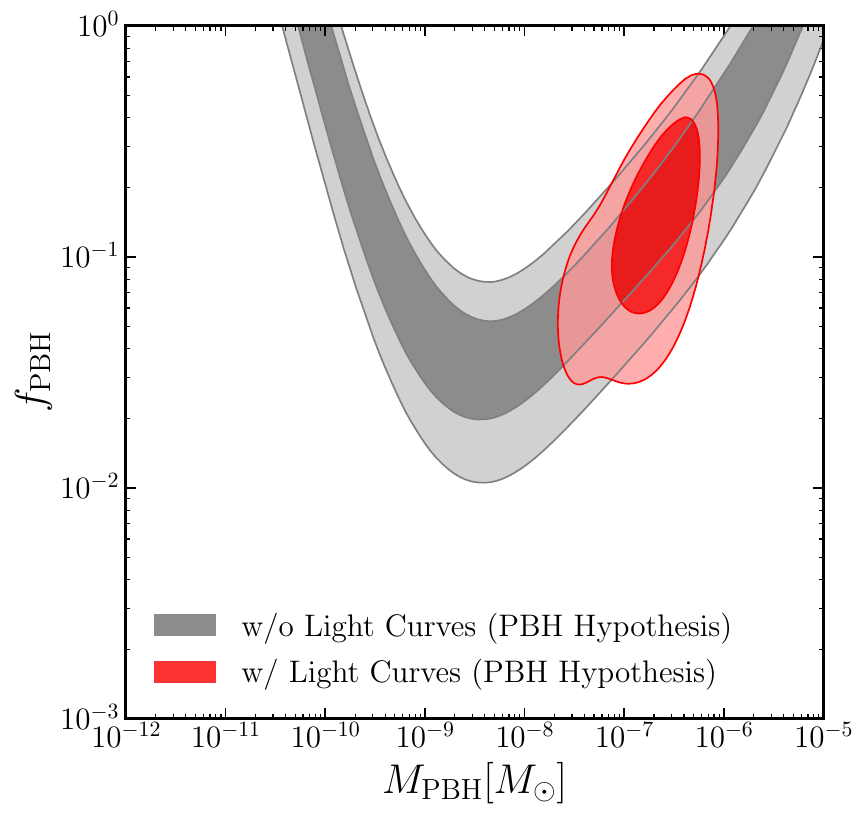}
    \includegraphics[width=0.48\textwidth]{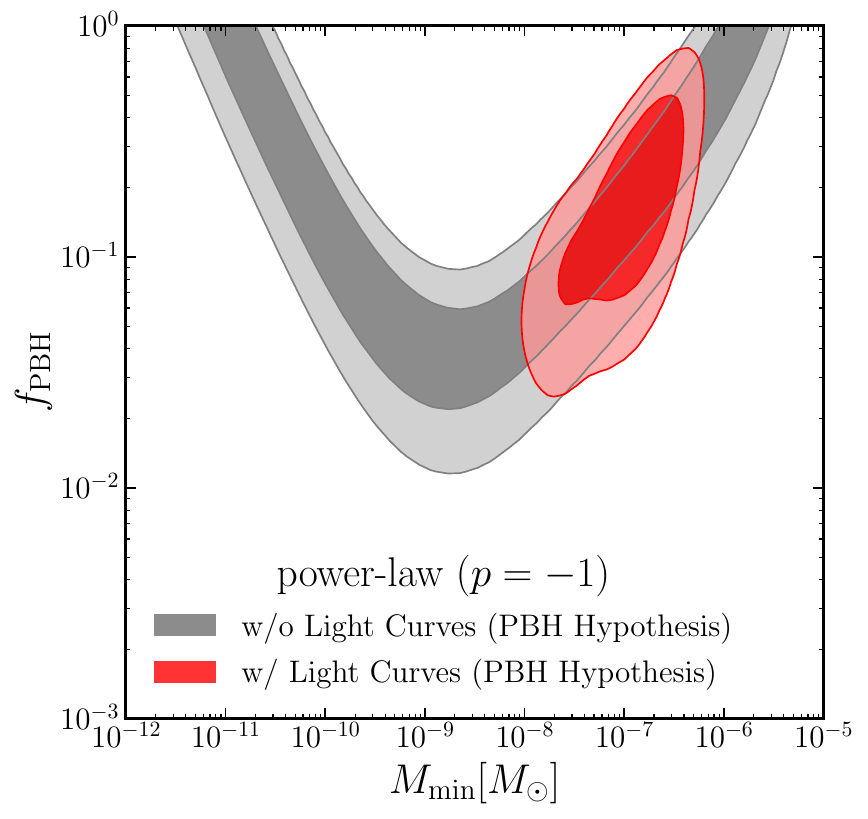}
    \caption{Similar to
    Fig.~\ref{fig:plausible}, but showing the results obtained using 
    only the 4 secure candidates (Table~\ref{tab:candidate-catalog}).
    A smaller abundance parameter is preferred compared to the previous 
    figure, as fewer events are attributed to PBH in this hypothesis. 
    The mass scale is almost the same though with larger 
    statistical uncertainty.    
    }
    \label{fig:plausible-secure}
\end{figure*}
Lastly in this section, we consider a scenario that lies 
between the previous two hypotheses; 
we assume that only the secure microlensing events defined in 
Section~\ref{ssec:candidate-secure-catalog}
are caused by PBHs.
The result is shown in Fig.~\ref{fig:plausible-secure}.
Because the number of the events used in the light-curve likelihood 
(Eq.~(\ref{eq:lc-like})) is smaller than the previous section, 
the constraint on the mass scale is weakened. 
Also, the uncertainty on the abundance parameter $f_{\rm PBH}$ 
is increased because of the smaller number of events used. 
As shown in Fig.~\ref{fig:lclike-profile}, the typical mass scales 
indicated by the light-curve likelihoods are similar 
in the candidate and secure catalogs; therefore, the central mass scale 
does not change from that shown in Fig.~\ref{fig:plausible}. 
However, the number of the events is smaller under this hypothesis 
than under the previous ones, and as a consequence 
the data prefer a smaller $f_{\rm PBH}$.

\section{Discussion and Conclusion}
\label{sec:conclusion}
\begin{figure*}
    \includegraphics[width=\textwidth]{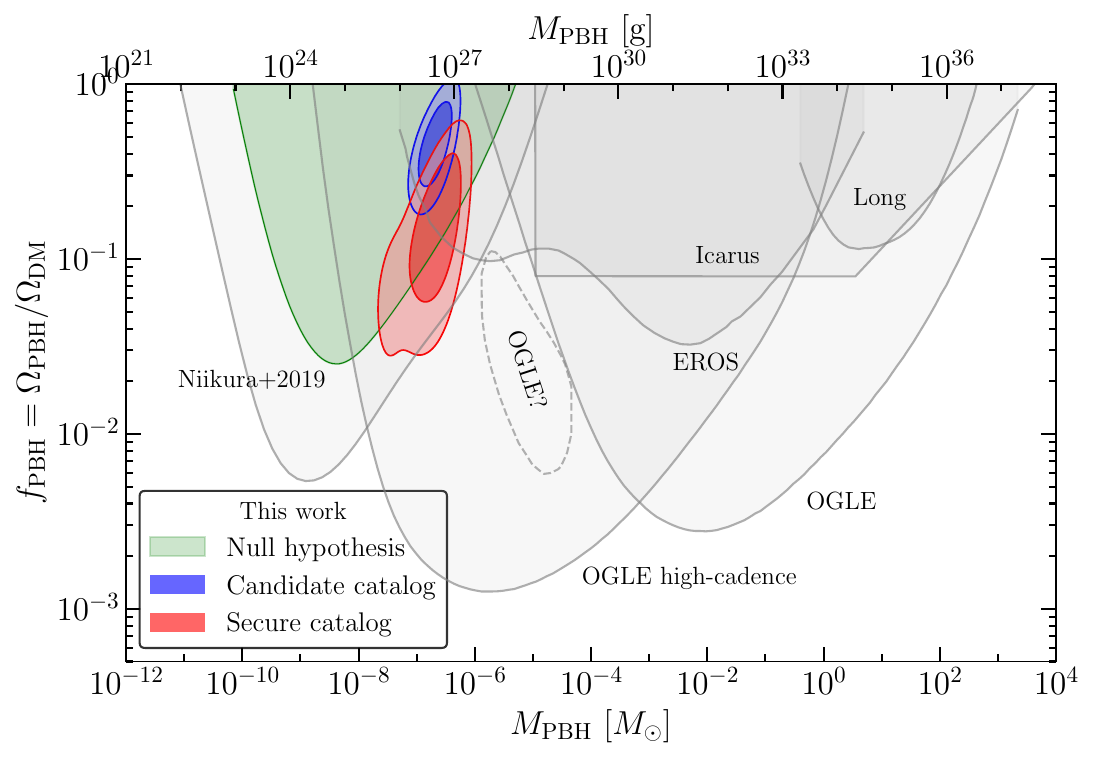}
    \caption{Comparison of the results in this paper with other results.
    The solid lines show the 95\% C.L. upper limits on $f_{\rm PBH}$ 
    at each PBH mass scale for a monochromatic mass function,
    while the dashed lines show the 95\% C.L. allowed regions.
    The upper limits from OGLE experiments denoted by ``OGLE'' 
    and ``OGLE high-cadence'' at high and low mass scales are from 
    \citet{Mroz.Ratajczak.2024, Mroz.Ratajczak.20248} 
    and \citet{Mroz.Mroz.2024} respectively.
    The allowed region
    denoted by ``OGLE?'' is from \citet{Niikura.Masaki.2019} 
    derived using the 6 short-time scale candidate events in OGLE experiment.
    The upper limit denoted by ``EROS'' is from \citet{Tisserand.Zylberajch.2006}.
    The upper limit denoted by ``Icarus'' is from \citet{Oguri.Broadhurst.2017}.
    The upper limit denoted by ``Long'' is from \citet{Blaineau.Tisserand.2022}.
    We used \citet{Kavanagh.Kavanagh} to plot 
    the upper limits from other papers.}
    \label{fig:comparison}
\end{figure*}

\subsection{Interpretation of Microlensing Candidates}
\label{sec:interpretation}

We compare our results obtained with the HSC data to those 
from other major (micro)lensing surveys and experiments. 
Fig.~\ref{fig:comparison} summarizes the constraints 
on the PBH abundance, $f_{\mathrm{PBH}}$, 
assuming a monochromatic PBH mass function. 
The solid lines represent upper limits derived under the null hypothesis 
from various microlensing searches.
The dashed lines indicate allowed regions of the PBH parameters, assuming 
that microlensing candidates are genuine,
as well as the region indicated by 5-year OGLE data (denoted as ``OGLE?'') \citep{Mroz.Pawlak.2017,Niikura.Masaki.2019}.

First, we compare our null-hypothesis upper limits with those obtained 
in \citet{Niikura.Chiba.2019}, which were derived using an earlier HSC 
dataset 
based on the 2014 data alone (subset of the dataset used in this work). 
Although both studies are based on HSC observations of M31,
our analysis incorporates a significantly expanded dataset, 
combining the 2014, 2017, and 2020 observations, 
as well as a revised treatment of the detection efficiency. 
Despite the increased data volume, our resulting upper limits are 
less stringent than those reported in \citet{Niikura.Chiba.2019}.
This difference arises primarily from
the fact that we include finite-source-size effects in 
the efficiency estimation.
While \citet{Niikura.Chiba.2019} injected only point-source microlensing events, 
leading to a relatively high detection efficiency, our analysis 
includes finite-source events in the injection process. 
As a consequence, the overall detection efficiency is reduced, 
which lowers the expected number of events and results in relaxed upper 
limits under the null hypothesis. 
This difference reflects not a loss of sensitivity in the data themselves, 
but rather a more realistic treatment of microlensing signals 
in the presence of finite source effects.

In contrast, the allowed region of PBH parameter space
derived under the PBH microlensing hypothesis 
is consistent with most existing constraints shown 
in Fig.~\ref{fig:comparison}, owing to differences in the mass ranges 
to which each experiment is most sensitive. 
However, 
it is 
in clear tension with the upper limits reported by 
the OGLE high-cadence microlensing result by \citet{Mroz.Mroz.2024}. 
Several possible explanations for this discrepancy can be considered.

\begin{itemize}
    \item First, it is possible that the candidate microlensing events
    identified 
    in this work are false positives, in which case the PBH hypothesis 
    would be disfavored. 
    The HSC observations are conducted in a single photometric band, 
    which prevents us from performing chromatic tests that could help 
    distinguish microlensing from intrinsic stellar variability. 
    In addition, because our analysis relies on pixel lensing, 
    the source stars are not resolved, and their intrinsic properties 
    cannot be directly constrained. 
    If such information were available, source colors or absolute magnitudes 
    could be used to infer distances and stellar types, providing a means 
    to identify contaminating variable stars. 
    While we cannot exclude this possibility, a detailed investigation 
    of source properties is beyond the scope of this paper.

    \item Second, the detection efficiency may be underestimated. 
    In our analysis, the efficiency is assumed to be determined by 
    a single representative parameter combination,
    $\hat{t} = 2 t_E \sqrt{u_T(\rho)^2 - u_0^2}$
    which effectively compresses the full, high-dimensional parameter space 
    into a one-dimensional description. 
    However, it is conceivable that the efficiency depends on 
    the full set of microlensing parameters in a non-trivial manner. 
    In particular, there may exist regions of parameter space where 
    the efficiency drops sharply, and events in such regions 
    would be systematically missed. 
    If these regions are improperly marginalized over, 
    the resulting efficiency could be misestimated. 
    A fully general treatment would require simulations spanning 
    the entire high-dimensional parameter space, 
    which is computationally prohibitive for the present study. 
    Nevertheless, it may be possible in future work to construct 
    an efficient emulator of the detection efficiency 
    using machine-learning techniques, 
    recasting the problem as a Bernoulli probability estimation task 
    that can be trained with a limited number of simulations. 
    Such an approach lies beyond the scope of this paper.

    \item Third, the OGLE high-cadence survey reports two remaining 
    candidate events, one of which is attributed to 
    a flare from a dwarf star, 
    while the other is interpreted as being caused 
    by a lens in the Milky Way disk.
    If either of these events were instead due to PBH microlensing, 
    the corresponding OGLE high-cadence upper limits would be weakened. 
    However, even under the extreme assumption that 
    both events are PBH-induced, 
    the resulting constraints would still be insufficient to 
    reconcile the discrepancy with the HSC results, 
    which differ at the level of nearly two orders of magnitude.

    \item Fourth, it is possible that the candidate 
    events detected in this work are all due to lenses 
    in the Milky Way disk rather than PBHs. 
    Because the line of sight toward M31 lies at a 
    lower Galactic latitude than those toward the LMC or SMC, 
    contamination from disk lenses could 
    in principle be more significant than 
    in the OGLE high-cadence observations. 
    To assess this possibility, we estimated the expected number of 
    microlensing events due to Milky Way disk lenses using a 
    Galactic disk model in \citet{Han.Gould.2003}. 
    We find that the expected contribution is extremely small, 
    e.g $N_{\mathrm{exp}}^{\mathrm{disk}} \simeq 10^{-6}$ for 
    2014-11-24 night, 
    and therefore negligible for the present analysis. 
\end{itemize}

In summary, it is difficult to attribute the tension 
between the HSC allowed
region and the OGLE high-cadence 
upper limits to any single effect. 
The OGLE high-cadence survey reports $\mathcal{O}(10^3)$ expected events 
but detects no more than two candidates, 
implying a discrepancy with the HSC results at the level of approximately 
two orders of magnitude. 
Resolving this tension will likely require a combination of improved 
control of detection efficiencies, better characterization of contaminating 
populations, and independent datasets with complementary systematics.

\subsection{Limitations and Future Improvements}

While the present analysis demonstrates the capability of imaging data 
to detect short-timescale microlensing events toward M31, several limitations remain, 
which also point to clear directions for future improvements.

First, the current data set is based on observations 
in a single photometric band (the $r$ band). 
Microlensing is, in principle, an achromatic phenomenon, 
as the gravitational deflection of light does not depend on wavelength 
in the geometric optics regime. 
Therefore, observing consistent magnification signals in multiple bands provides 
a powerful discriminator between genuine microlensing events and intrinsic stellar variability. 
Although chromatic effects may arise in the wave-optics regime or due to 
finite-source effects combined with limb darkening, multi-band observations 
would significantly enhance the robustness of event classification. 
Future surveys with multi-band time-series imaging will allow us 
to explicitly test achromaticity and reduce contamination from variable stars.

Second, the temporal coverage of observations plays a critical role 
in the detectability of microlensing events. 
In the current strategy, observations are often limited to partial nights, 
which restricts the effective time baseline. 
Extending the coverage to full-night observations would substantially improve 
sensitivity, even for short-timescale events. 
This is because the expected number of detectable events increases more than 
linearly with the observation time, as longer continuous coverage improves 
the overall observing time as well as the detection efficiency. 
Therefore, having one night observing allocation would be helpful 
to maximize the sensitivity to the microlensing events.

Third, the false positive rate remains an important concern. 
Although the current selection criteria are designed to suppress 
known classes of contaminants, 
such as variable stars and subtraction artifacts, 
residual false positives may still be present. 
A more systematic characterization of the false positive rate, 
potentially using dedicated control samples or injection–recovery tests on real data, 
will be necessary to fully quantify the purity of the event sample. 
Such studies will also inform the optimal balance between 
completeness and reliability in future analyses.

Fourth, the detection efficiency is currently characterized 
using a limited set of representative parameters, typically focusing 
on a single timescale variable $\hat{t}$. 
In reality, the efficiency depends on multiple correlated parameters, 
including event timescale, 
impact parameter, finite-source size parameter. 
A more accurate efficiency model that accounts for this multi-dimensional parameter dependence 
would lead to a more precise inference of the underlying microlensing event rate and, 
consequently, tighter constraints on compact-object populations. 
Developing such a method is an important goal for future work.

Finally, image quality, particularly the seeing condition, plays 
a crucial role in our analysis. 
Since the detection relies on image subtraction, high-quality reference images 
with good seeing are essential for minimizing residuals and 
improving sensitivity to faint and short-duration events. 
Future observations that prioritize excellent seeing for reference image construction, 
as well as improved point-spread-function modeling, 
will further enhance the performance of difference imaging.

In summary, while the present study establishes a solid foundation 
for microlensing searches toward M31, incorporating multi-band observations, 
longer and more continuous temporal coverage, improved false-positive control, 
more sophisticated efficiency modeling, and higher-quality reference imaging 
will be essential for fully exploiting the potential of this approach in future surveys.

Finally, future next-generation surveys provide useful prospects for 
short-timescale microlensing searches based on difference imaging, 
as adopted in this study, from different perspectives. 
The Nancy Grace Roman Space Telescope has a highly stable and 
well-characterized point-spread function as a space-based telescope, 
which is expected to reduce systematic errors in difference imaging 
analyses in crowded stellar fields. 
This property will allow more 
precise measurements of microlensing light curves, including short-timescale 
events and finite-source effects. On the other hand, 
the Vera C. Rubin Observatory (LSST) will provide wide-field, 
long-term time-series observations, which are expected to increase 
the microlensing event sample and to reduce statistical uncertainties 
in the event rate estimation. In addition, its high observing cadence 
and long temporal baseline will provide important information for 
evaluating the sensitivity to short-timescale events. 
Considering these characteristics, next-generation surveys are 
expected to provide useful data sets for future applications and 
tests of the method presented in this work.

\begin{acknowledgments}

We thank Takayuki Ohgami for helping us analyzing the image data at the early stage of the paper.
We thank Tsuyoshi Terai, Mirko Simunovic, and Takuya Fujiyoshi for the service as Support Astronomers.
We thank Hiroko Niikura for providing the previous analysis pipeline.
SS thank Satoshi Toki for a useful discussion about the hierarchical Bayesian inference of the PBH mass function, and providing us with the code to 
calculate the event rate by disk stars.
We thank Alex Kusenko for the useful comments on the theoretical PBH model.
This work was supported in part by JSPS KAKENHI Grant Number 23KJ0747,
and by World Premier International Research Center Initiative (WPI Initiative), MEXT,
Japan.

The Hyper Suprime-Cam (HSC) collaboration includes the astronomical communities of Japan and Taiwan, and Princeton University. The HSC instrumentation and software were developed by the National Astronomical Observatory of Japan (NAOJ), the Kavli Institute for the Physics and Mathematics of the Universe (Kavli IPMU), the University of Tokyo, the High Energy Accelerator Research Organization (KEK), the Academia Sinica Institute for Astronomy and Astrophysics in Taiwan (ASIAA), and Princeton University. Funding was contributed by the FIRST program from the Japanese Cabinet Office, the Ministry of Education, Culture, Sports, Science and Technology (MEXT), the Japan Society for the Promotion of Science (JSPS), Japan Science and Technology Agency (JST), the Toray Science Foundation, NAOJ, Kavli IPMU, KEK, ASIAA, and Princeton University. 

This paper makes use of software developed for Vera C. Rubin Observatory. We thank the Rubin Observatory for making their code available as free software at http://pipelines.lsst.io/.

This paper is based on data collected at the Subaru Telescope and retrieved from the HSC data archive system, which is operated by the Subaru Telescope and Astronomy Data Center (ADC) at NAOJ. Data analysis was in part carried out with the cooperation of Center for Computational Astrophysics (CfCA), NAOJ. We are honored and grateful for the opportunity of observing the Universe from Maunakea, which has the cultural, historical and natural significance in Hawaii. 

The Pan-STARRS1 Surveys (PS1) and the PS1 public science archive have been made possible through contributions by the Institute for Astronomy, the University of Hawaii, the Pan-STARRS Project Office, the Max Planck Society and its participating institutes, the Max Planck Institute for Astronomy, Heidelberg, and the Max Planck Institute for Extraterrestrial Physics, Garching, The Johns Hopkins University, Durham University, the University of Edinburgh, the Queen’s University Belfast, the Harvard-Smithsonian Center for Astrophysics, the Las Cumbres Observatory Global Telescope Network Incorporated, the National Central University of Taiwan, the Space Telescope Science Institute, the National Aeronautics and Space Administration under grant No. NNX08AR22G issued through the Planetary Science Division of the NASA Science Mission Directorate, the National Science Foundation grant No. AST-1238877, the University of Maryland, Eotvos Lorand University (ELTE), the Los Alamos National Laboratory, and the Gordon and Betty Moore Foundation.
\end{acknowledgments}

\bibliography{refs_ss}

\onecolumngrid

\appendix
\section{Impact of using lower quality reference image}
\label{apdx:bad-seeing-reference}
\begin{figure*}
    \includegraphics[width=0.45\textwidth]{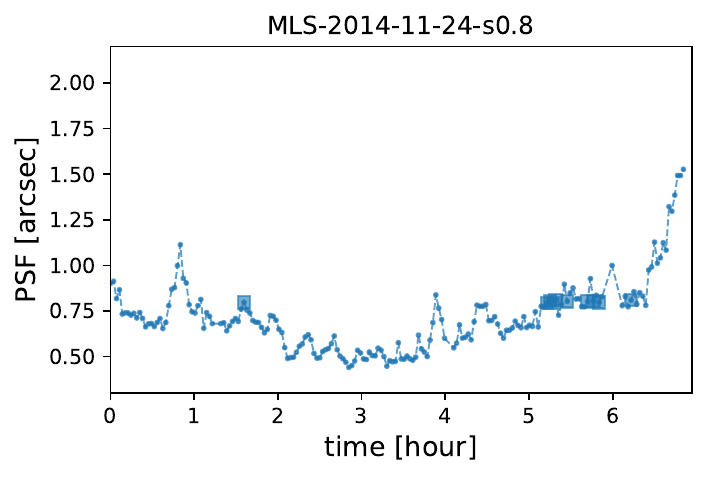}
    \caption{Same figure as Fig.~\ref{fig:seeing}, but with the square data points indicating the 10 images with seeing size $\sim0.8\arcsec$ that are used for making reference images in Appendix~\ref{apdx:bad-seeing-reference}.}
    \label{fig:seeing-s08}
\end{figure*}
\begin{table*}
    \centering
    \caption{
    The selection process when we use relatively lower-quality images 
    for the reference image. The row names as ``2014-11-24'' is the 
    fiducial analysis using the best images with smallest seeing size,
    and identical to the row shown in Table.~\ref{tab:selection-result}.
    The bottom row named as ``2014-11-24-s0.8'' is still using the same 
    image data from 2014-11-24, but using the reference image is made by 
    coadding the 10 images whose seeing size is about $0.8\arcsec$.
    }
\begin{tabular}{lrrrrrrrrrrrrrrr}
\toprule
ID & master & bump & mlc2 & mlc2i & mlc2o & mlc2il & mlc2ir & mlc2ol & mlc2or & asym & scorr & scorrs & ntscale & amax & mlsig \\
\midrule
2014-11-24 & 7281 & 5139 & 1135 & 938 & 911 & 878 & 819 & 814 & 783 & 391 & 72 & 70 & 57 & 53 & 38 \\
2014-11-24-s0.8 & 5715 & 5337 & 1113 & 963 & 931 & 880 & 837 & 832 & 807 & 289 & 62 & 0 & 0 & 0 & 0 \\
\bottomrule
\end{tabular}
    \label{tab:selection-result-s08}
\end{table*}

In this appendix, we investigate the impact of using lower-quality images to construct 
the reference image. For comparison with the best-performing scenario, we use data from 
2014-11-24, which have the highest image quality and the smallest seeing values.

We select 10 images with seeing sizes of approximately $0.8\arcsec$ and create a 
lower-quality reference image by coadding them. The seeing of the images used for this 
experiment are indicated in Fig.~\ref{fig:seeing-s08}. We then obtain difference images 
by subtracting this lower-quality reference image from each individual image. Image-level 
detection and photometric measurements are performed on these difference images to produce  another master catalog. Finally, we apply the same selection criteria as in the main analysis described in 
Section~\ref{sec:data-analysis} to the master catalog to identify microlensing candidates. 
The results are summarized in Table~\ref{tab:selection-result-s08}, together with those 
from the fiducial analysis for ease of comparison.
After applying all selection criteria, no events remain. This result indicates that the 
high quality of the reference image is crucial for the efficient detection of microlensing 
events.

\section{Event rate and event number}\label{apdx:eventrate-model}
We start defining the differential event rate by following \citet{Niikura.Chiba.2019}. We define the differential eventrate with respect to four variables: the distance to the lens object $d$, the event time scale $\hat{t}$, the angle to enter the Einstein circle $\theta$ on the sky, and the physical source radius $R_{\rm s}$. We group the part of variables as $\bm{\theta}=\{\hat{t}, \theta, R_{\rm s}\}$ for the later purpose. 
The differential eventrate is 
\begin{align}
    \frac{\dd \Gamma}{\dd d \dd\bm{\theta}'} = 
    f(R_{\rm s})
    \sum_{h \in \text{M31, MW}}
    \frac{\rho_{{\rm DM}, h}(d)v_{{\rm c},h}^2}{M_{\rm PBH}}
    \left(\frac{v_{\rm r}}{v_{{\rm c},h}}\right)^4 
    \exp\left[-\frac{v_{\rm r}^2}{v_{{\rm c},h}^2}\right]
    \label{eq:differential-eventrate}
\end{align}
where $\rho_{{\rm DM}, h}$ and $v_{{\rm c},h}$ are the mass profile and the velocity dispersion of the $h$-th galaxy halo, which is either of Milky Way (MW) or M31 galaxy. The velocity realization of the lens object that belongs to $h$-th galaxy halo is defined as $v_{\rm r} = 2R_{\rm E}u_{\rm T}(\rho)\cos\theta/\hat{t}$ where $\rho=\theta_{\rm s}/\theta_{\rm E}=(R_{\rm s}/d_{\rm s})/(R_{\rm E}/d)$ is the source size radius in the unit of Einstein radius on the lens plane, and $u_{\rm T}$ is the minimum impact parameter within which the magnification becomes a certain threshold ($A=1.34$).
We introduced the distribution of the background source sizes $f_{\rm s}(R_{\rm s})$ with $R_{\rm s}$ a physical size of the source radius.
We use the source-size distribution estimated in \citet{Smyth.Guhathakurta.2019}, and normalize it such that $\int\dd R_{\rm s} f_{\rm s}(R_{\rm s})=1$.

To obtain the differential eventrate only for $\hat{t}$, we can integrate over all other variables,
\begin{align}
    \frac{{\rm d}\Gamma}{{\rm d}\hat{t}} = 
    \int_0^{d_{\rm s}}\dd d 
    \int_0^{\infty}\dd R_{\rm s}
    \int_{-\pi/2}^{\pi/2}\dd \theta
    \frac{\dd \Gamma}{\dd d \dd\bm{\theta}'}
\end{align}
The integral over $\theta$ can be analytically performed to reduce the expression to
\begin{align}
    \frac{{\rm d}\Gamma}{{\rm d}\hat{t}} = 
    \sum_{h}
    \int_0^{d_{\rm s}}\dd d 
    \int_0^{\infty}\dd R_{\rm s}
    f_{\rm s}(R_{\rm s})
    \frac{\rho_{{\rm DM},h}(d)v_{{\rm c},h}^2}{M_{\rm PBH}} 
    I\left( \frac{2R_{\rm E}u_{\rm T}}{v_{{\rm c},h}\hat{t}} \right)
\end{align}
$I(x)=\pi/2x^2e^{-x^2/2}[x^2 I_0(x^2/2) - (1+x^2)I_1(x^2/2)]$ with $I_{0,1}$ the first kind of modified Bessel function of zero and first order.

By integrating over the event time scale, accounting for the efficiency, we obtain the event rate for $n$-th night
\begin{align}
    \Gamma_{n}
    =\int_0^\infty {\rm d}\hat{t} \epsilon_{{\rm eff}, n}(\hat{t})
    \frac{{\rm d}\Gamma}{{\rm d}\hat{t}}
    \label{eq:eventrate}
\end{align}
where $\epsilon_{{\rm eff}, n}$ is the effective detection efficiency introduced in Eq.~(\ref{eq:effective-efficiency}), accounting for the magnitude dependence of the efficiency and the numbers of source stars in different magnitude bins.

The number of events for the $n$-th night is then obtained by
\begin{align}
    N_{{\rm exp}, n} = N_{{\rm tot}} T_{{\rm eff}, n} \Gamma_{n}.
    \label{eq:nexp-date}
\end{align}
The total number of the events for whole observations can be simply obtained by summing the expected number of events for each observation date; $N_{{\rm exp, tot}}=\sum_{n}N_{{\rm exp},n}$.

\section{Light curves}\label{apdx:light-curve}
\begin{figure*}
\centering
\begin{minipage}{0.48\textwidth}
  \includegraphics[width=\linewidth]{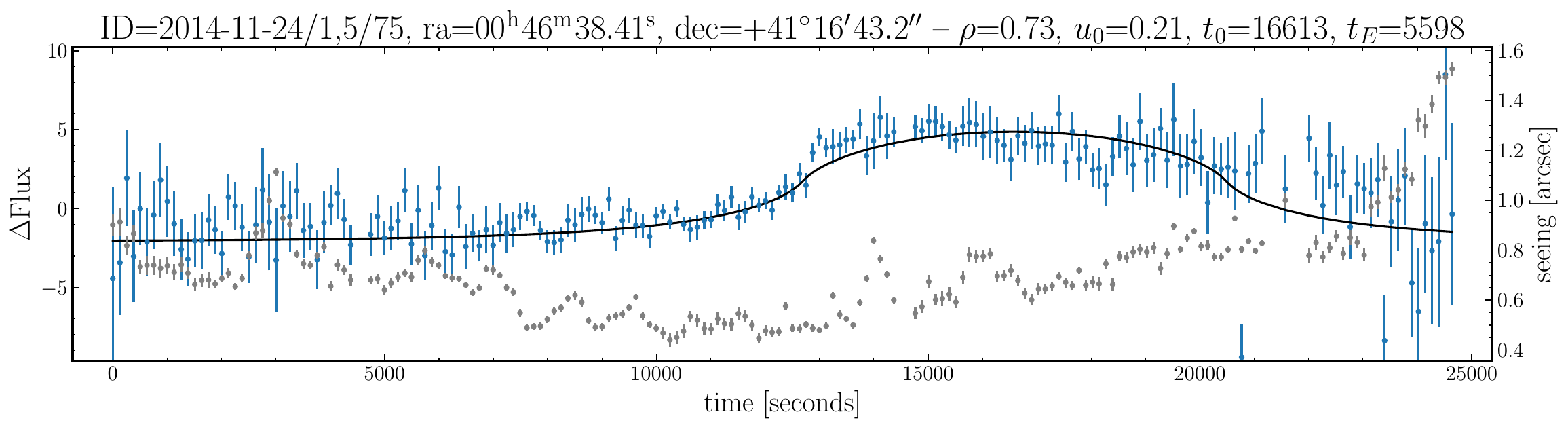}
  \includegraphics[width=\linewidth]{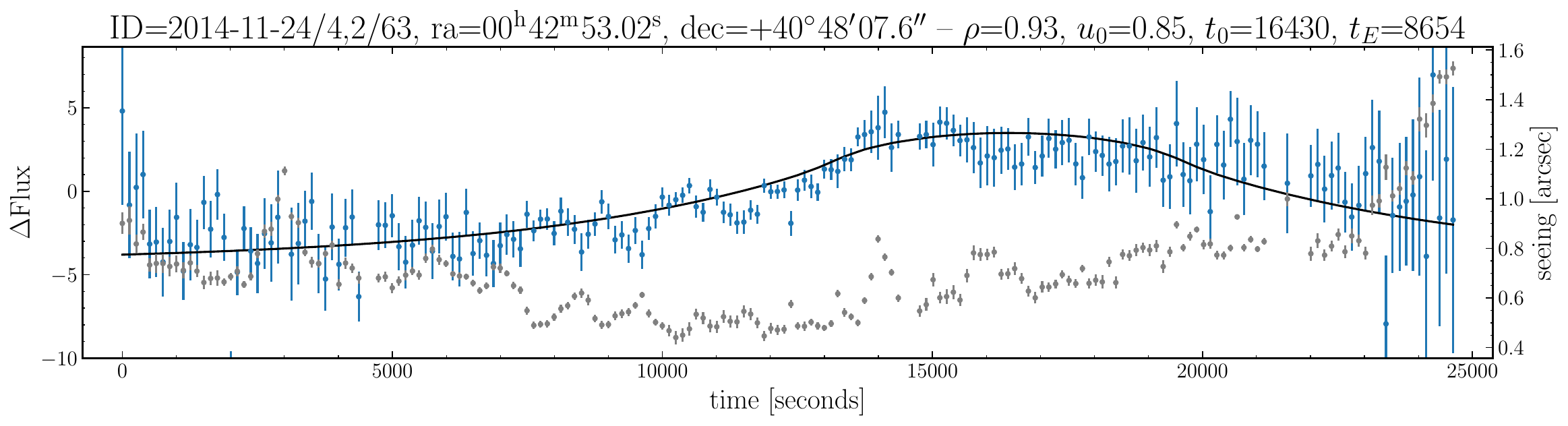}
  \includegraphics[width=\linewidth]{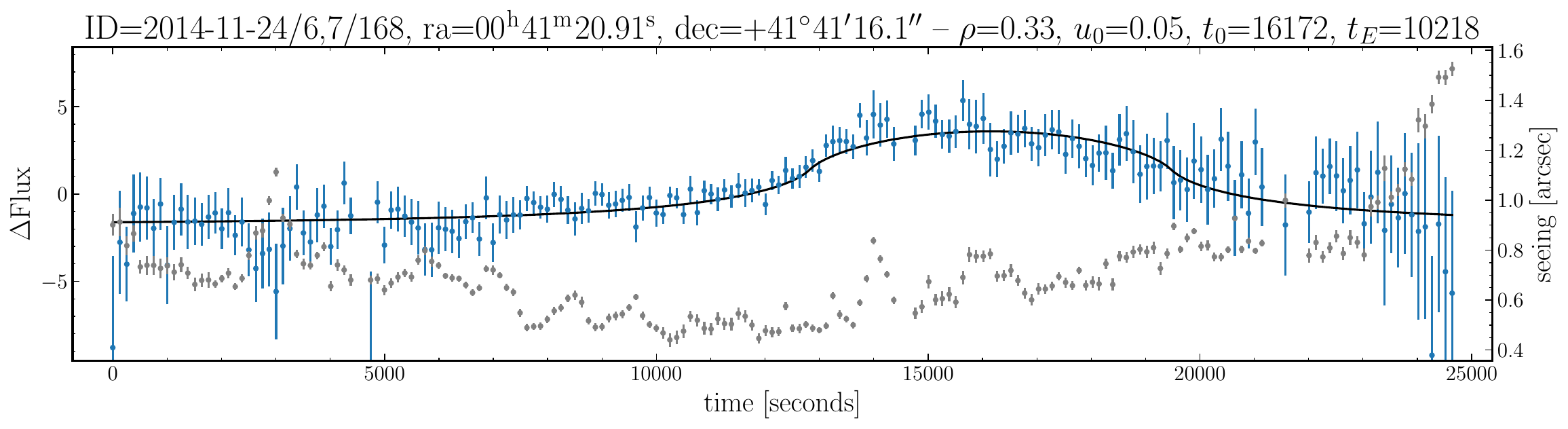}
  \includegraphics[width=\linewidth]{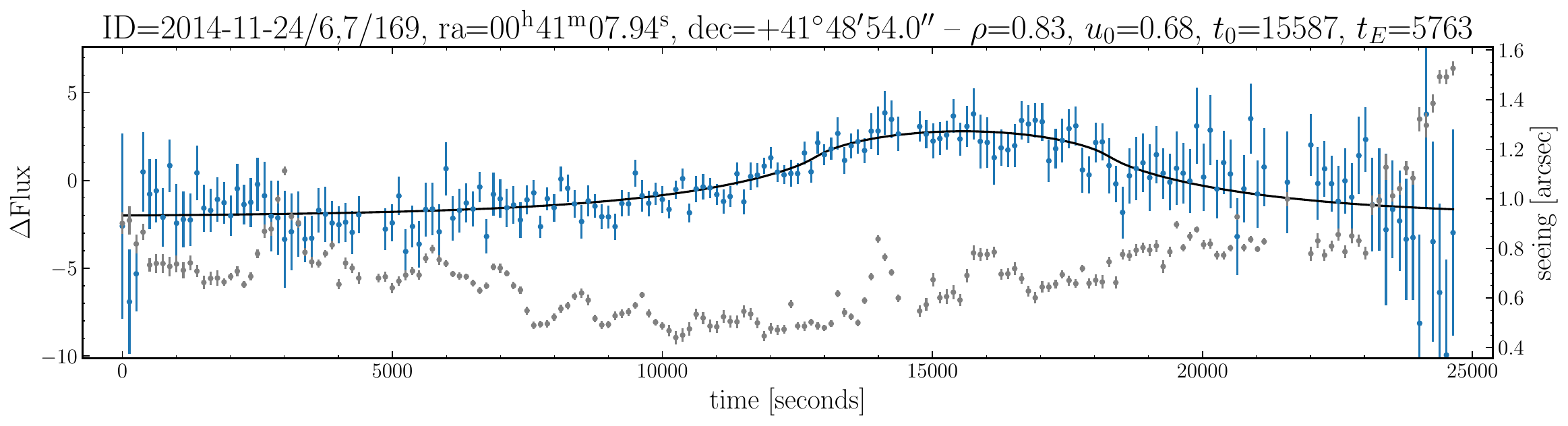}
  \includegraphics[width=\linewidth]{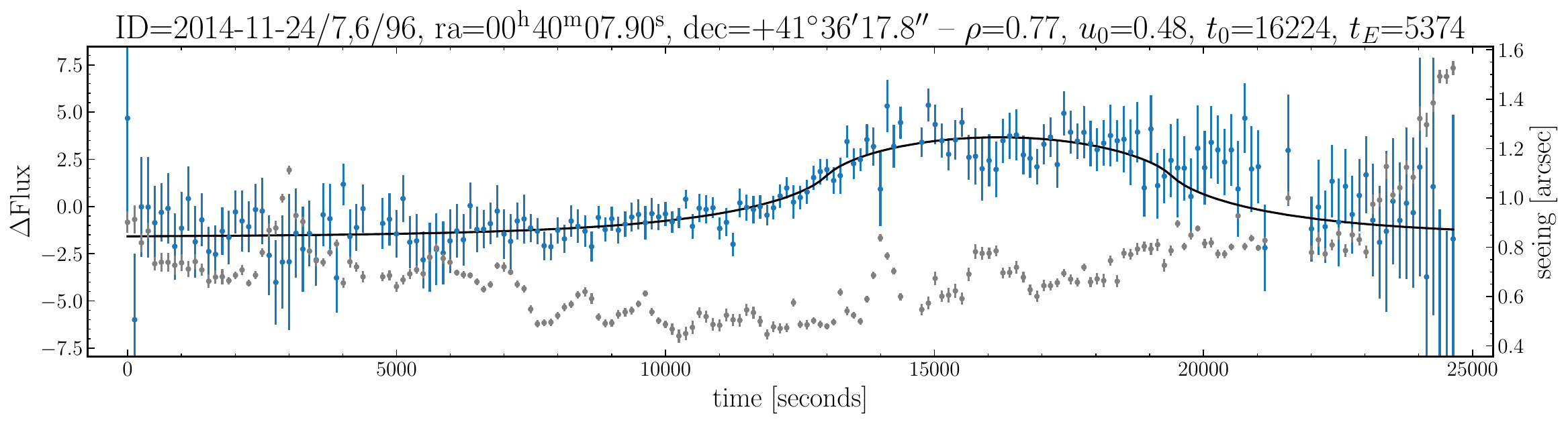}
  \includegraphics[width=\linewidth]{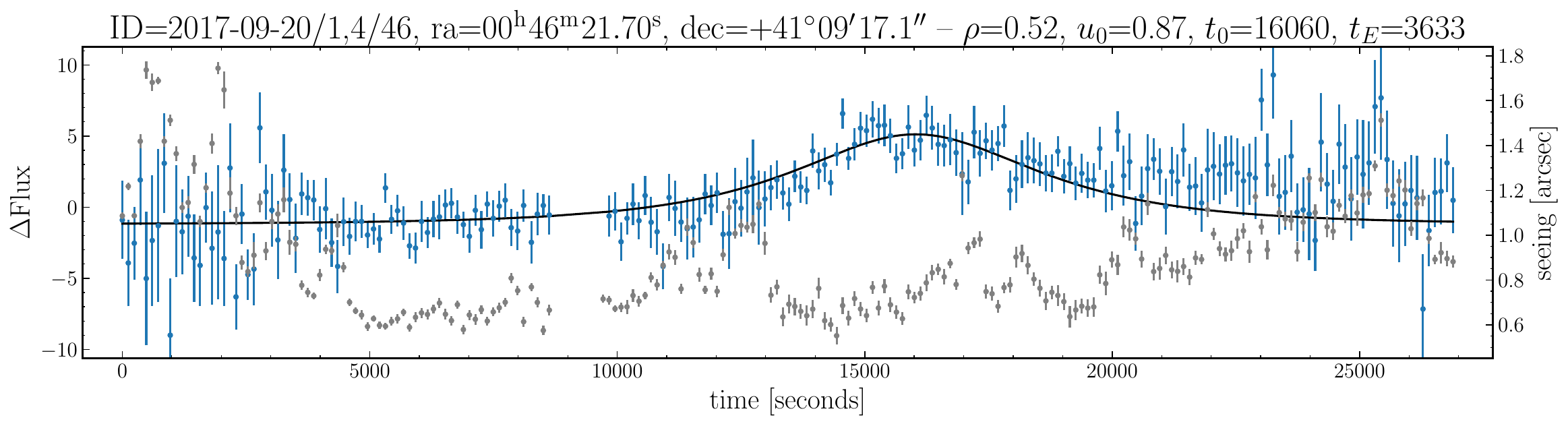}
\end{minipage}
\hfill
\begin{minipage}{0.48\textwidth}
  \includegraphics[width=\linewidth]{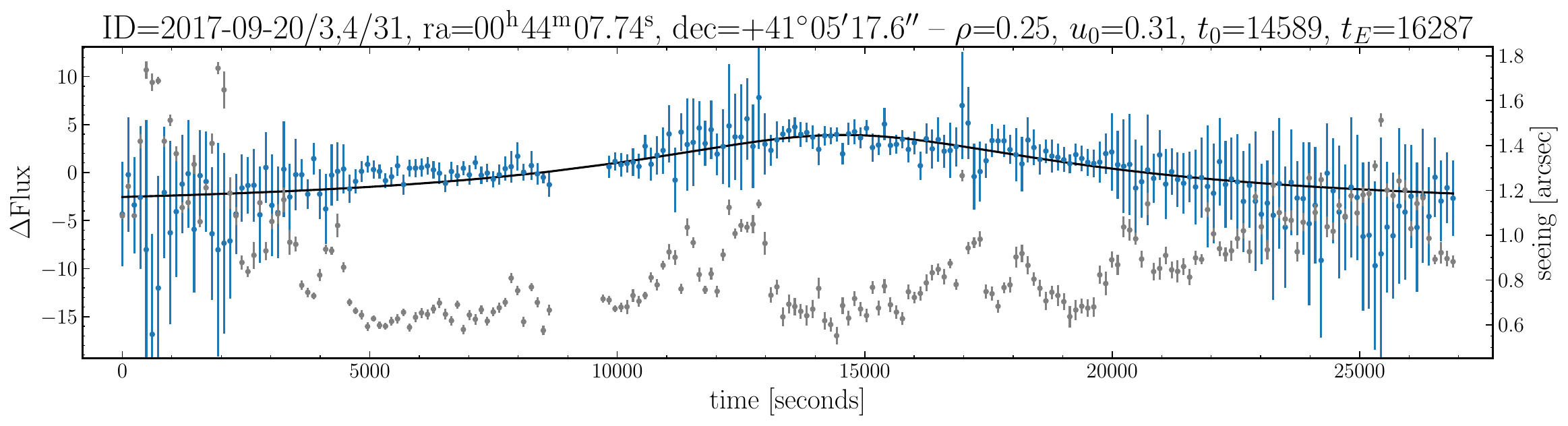}
  \includegraphics[width=\linewidth]{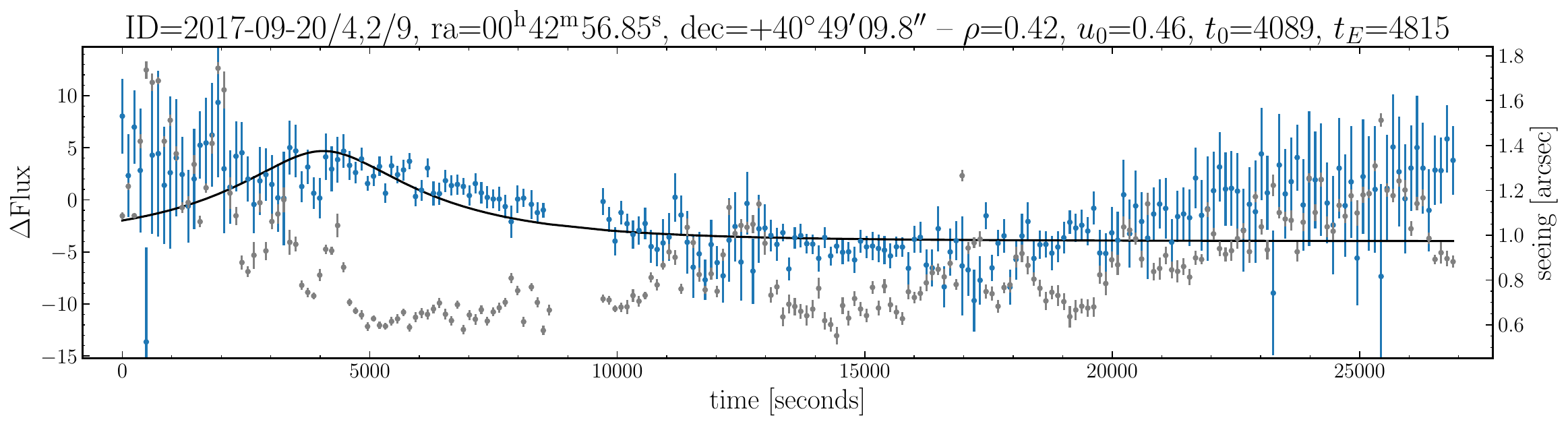}
  \includegraphics[width=\linewidth]{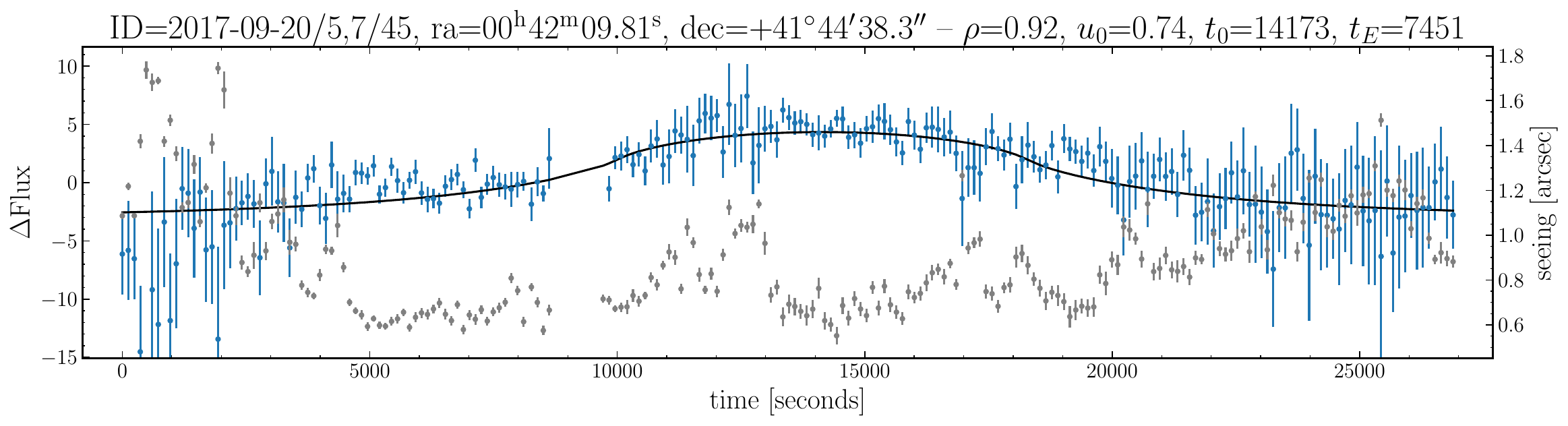}
  \includegraphics[width=\linewidth]{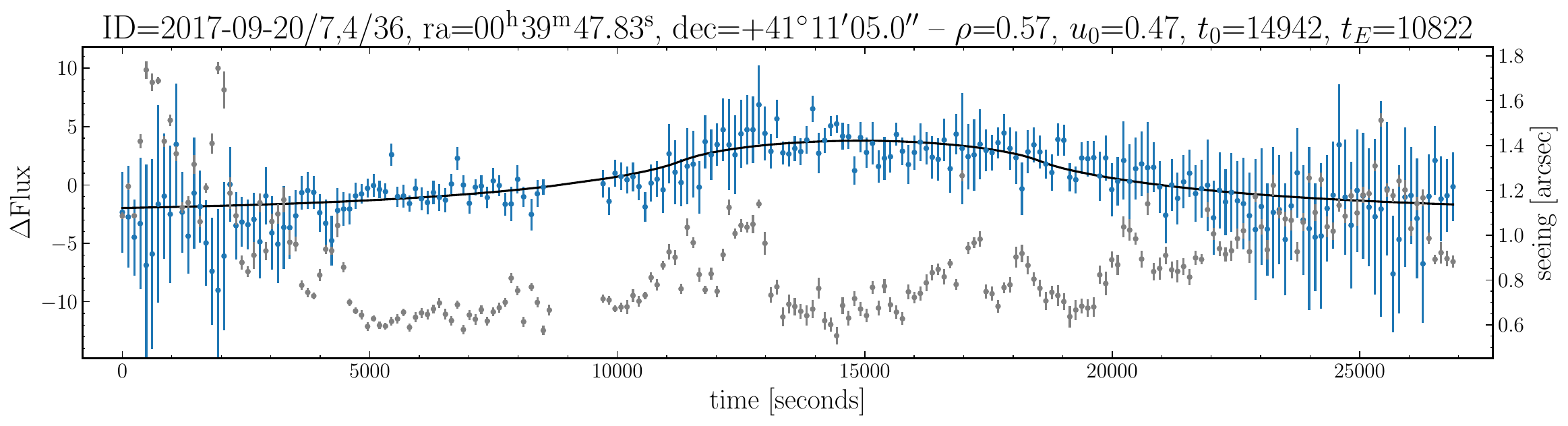}
  \includegraphics[width=\linewidth]{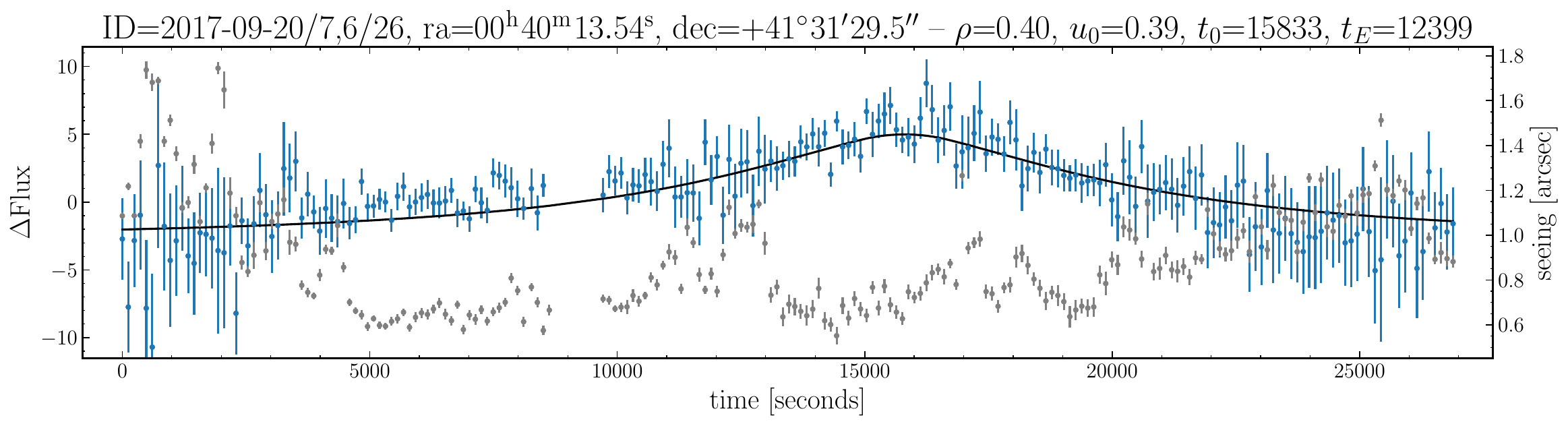}
  \includegraphics[width=\linewidth]{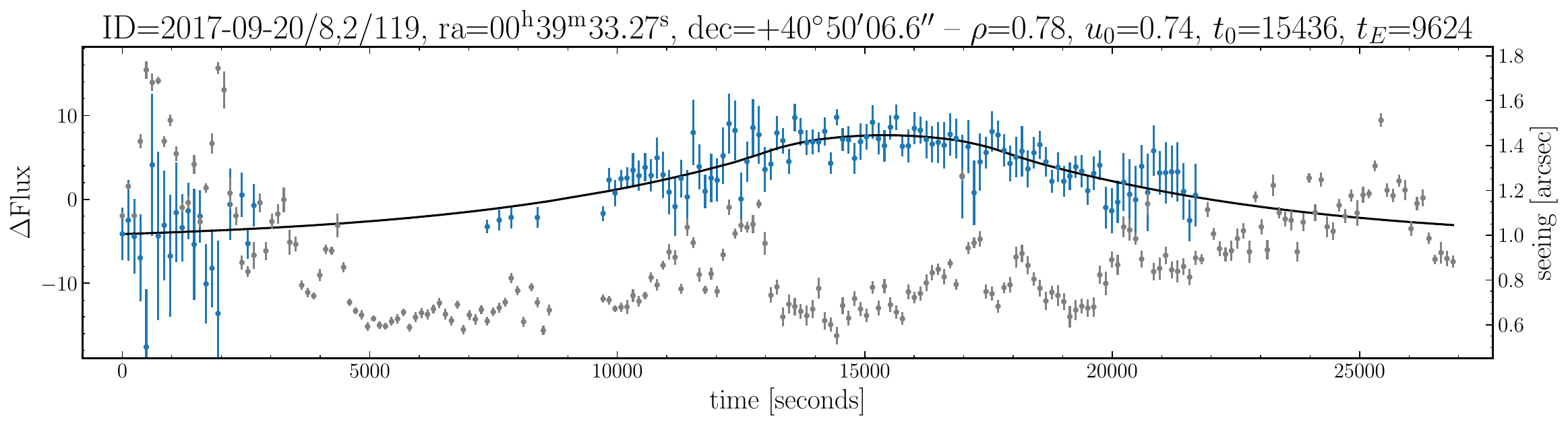}
\end{minipage}
\caption{Light curves of the 12 selected events in the candidate catalog the property of these events are summarized in Table~\ref{tab:candidate-catalog}. The blue data points are the flux data points obtained by the forced photometry on the detection point on the difference image. The black curve is the theoretical prediction by the microlensing PLFS model with the best-fit parameters of light curve fitting.}
\label{fig:lightcurve}
\end{figure*}
In this appendix, we present the light curves of the events in the candidate catalog selected from master catalog in Section~\ref{ssec:event-selection}. 
Fig.~\ref{fig:lightcurve} shows the light curves together with the theoretical prediction of PLFS model using the best-fit parameters of light curve fitting.

\section{Posteriors of light-curve fitting for selected events}
\label{apdx:lc-post}
\begin{figure*}
\includegraphics[width=0.3\textwidth]{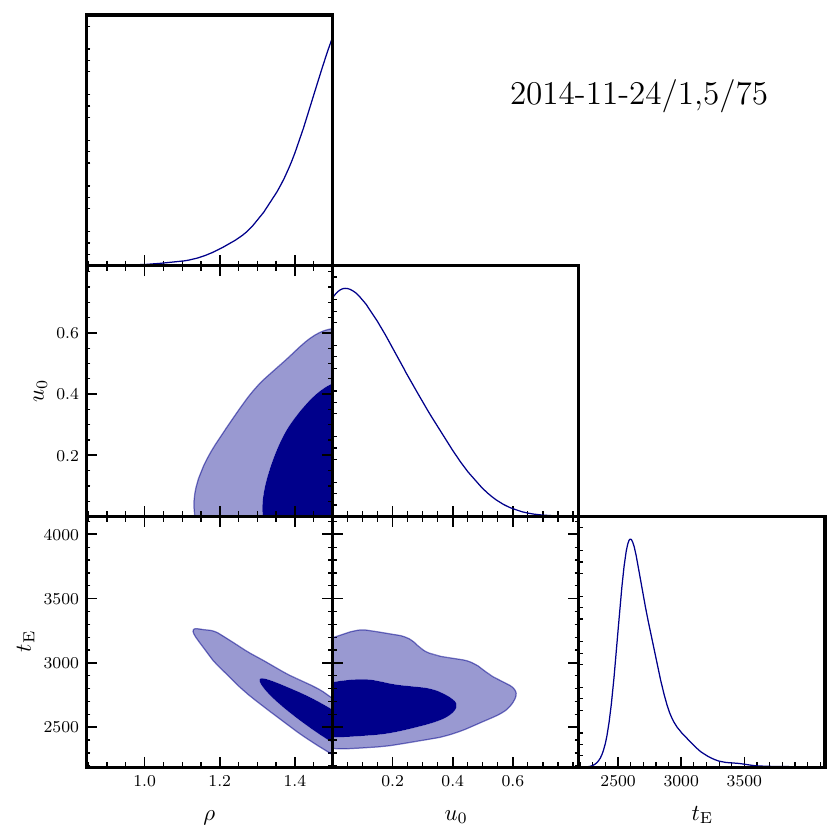}
\includegraphics[width=0.3\textwidth]{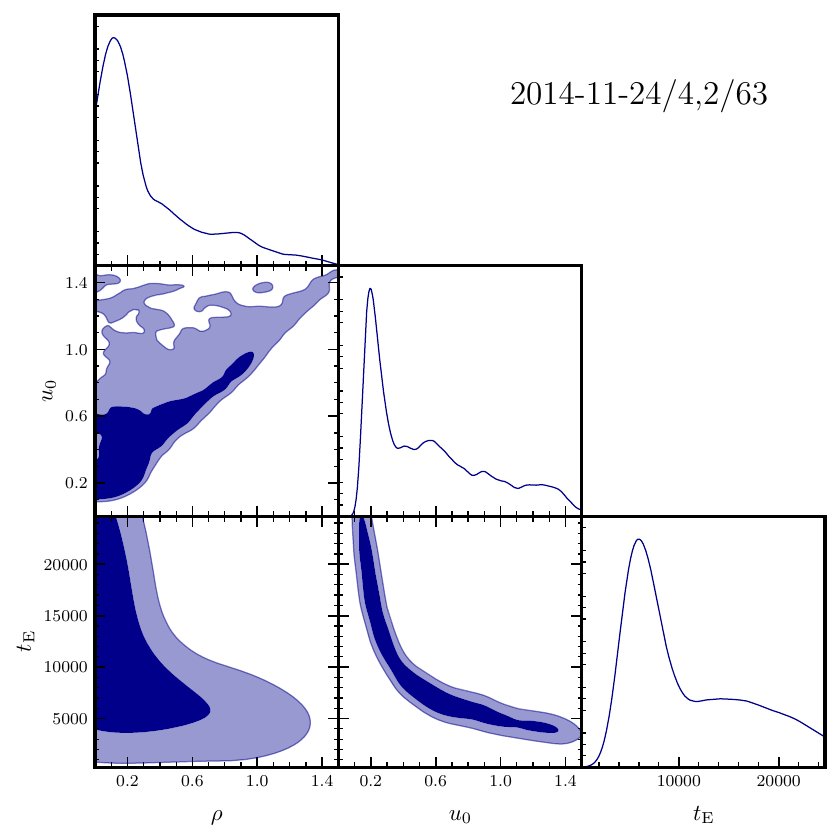}
\includegraphics[width=0.3\textwidth]{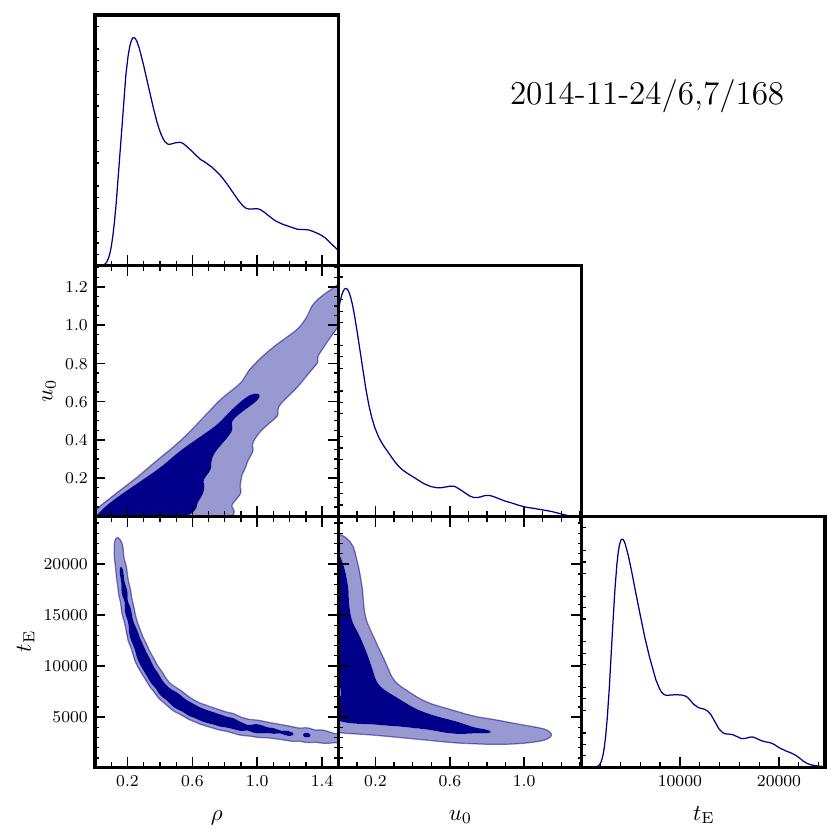}
\includegraphics[width=0.3\textwidth]{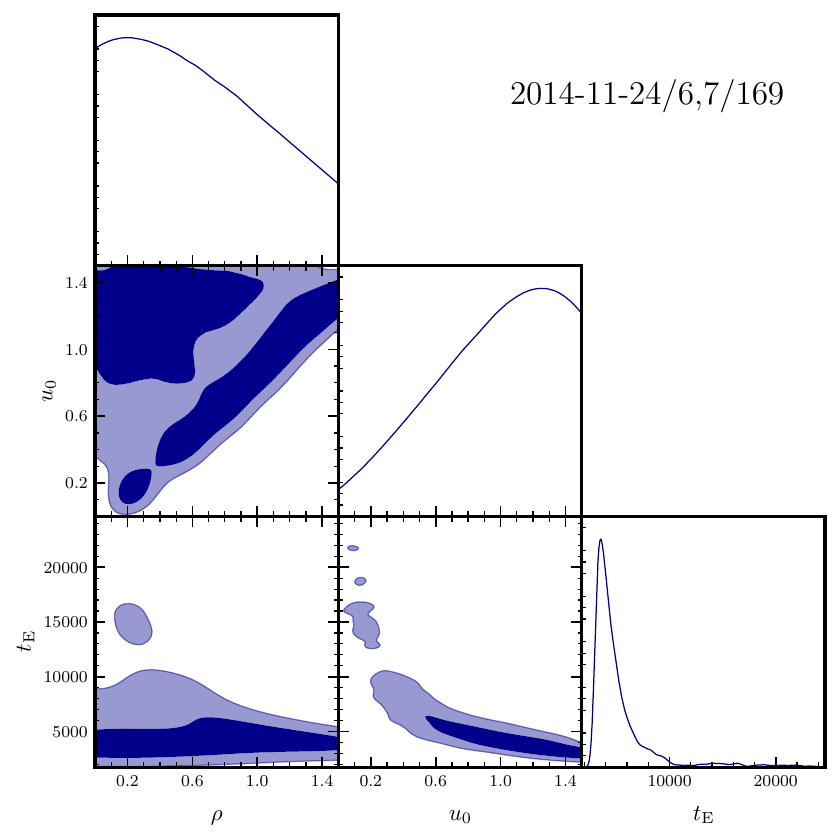}
\includegraphics[width=0.3\textwidth]{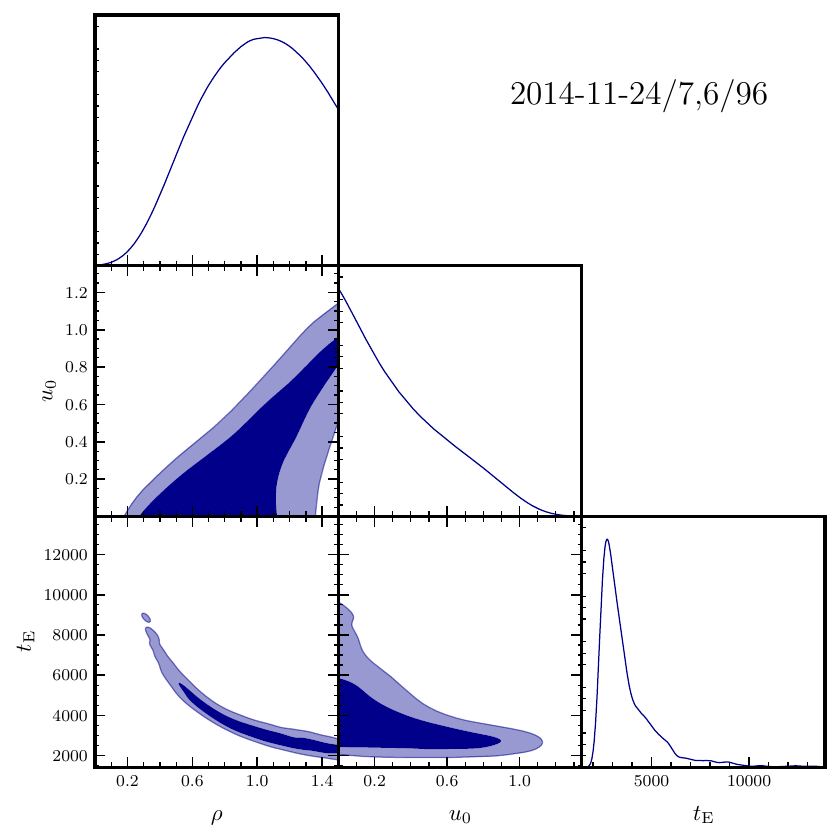}
\includegraphics[width=0.3\textwidth]{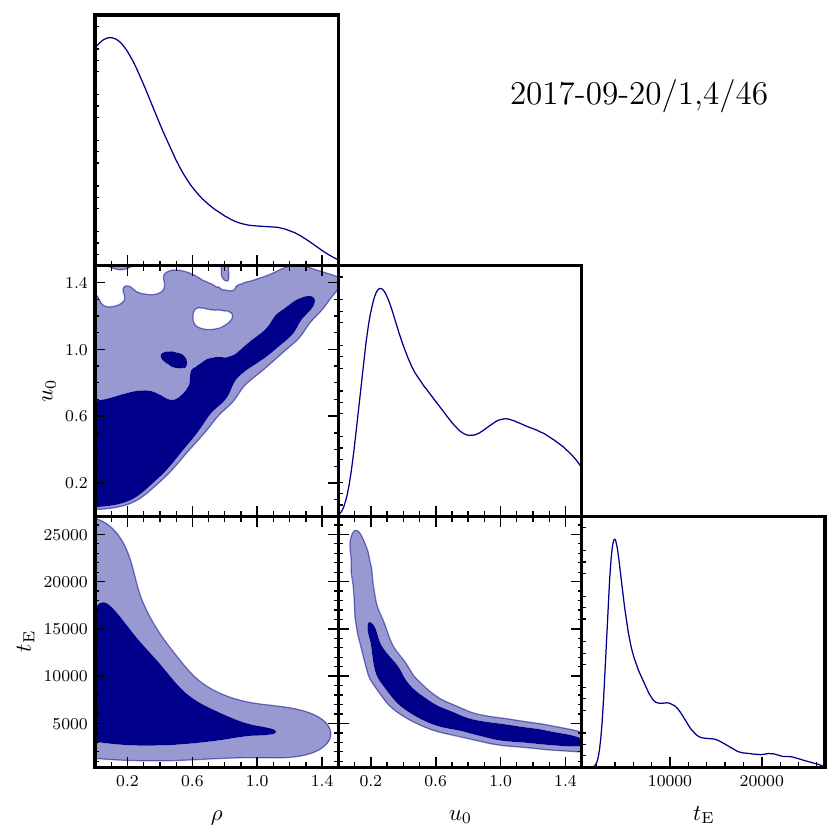}
\includegraphics[width=0.3\textwidth]{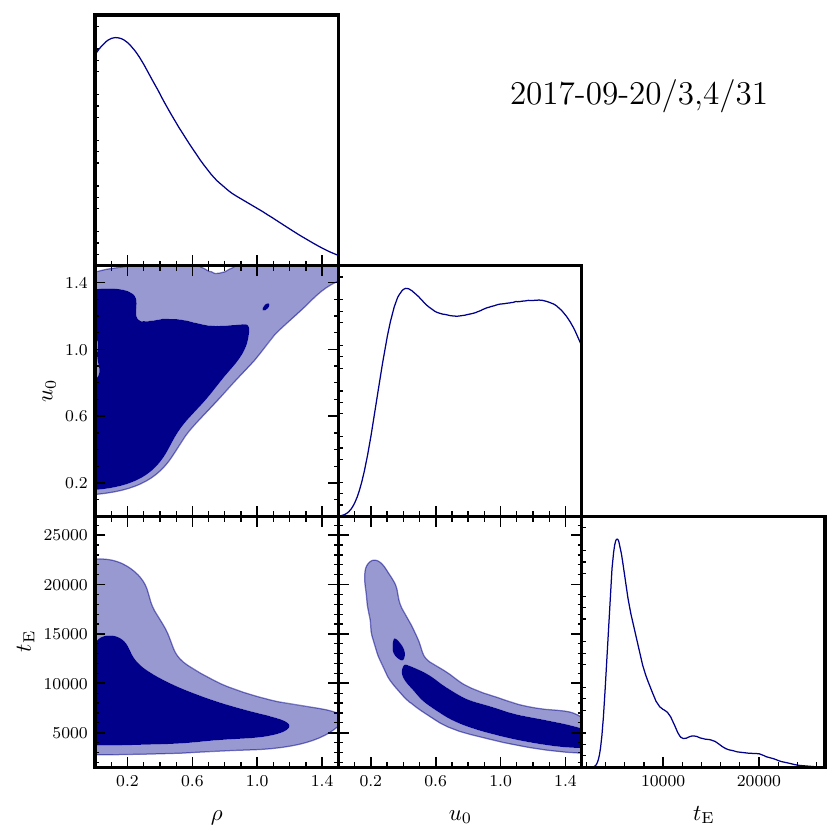}
\includegraphics[width=0.3\textwidth]{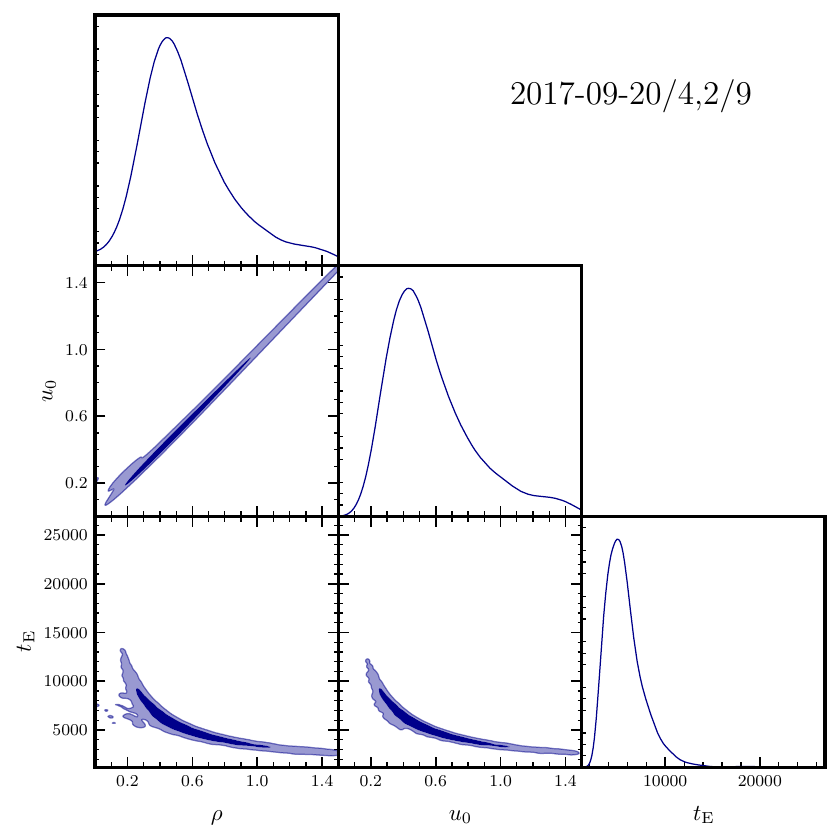}
\includegraphics[width=0.3\textwidth]{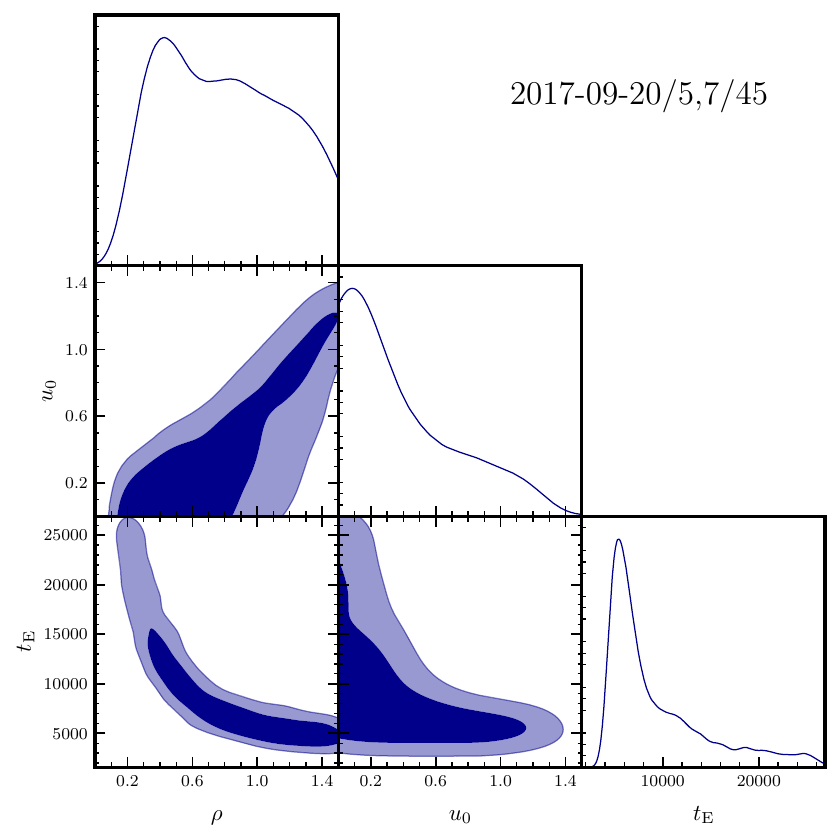}
\includegraphics[width=0.3\textwidth]{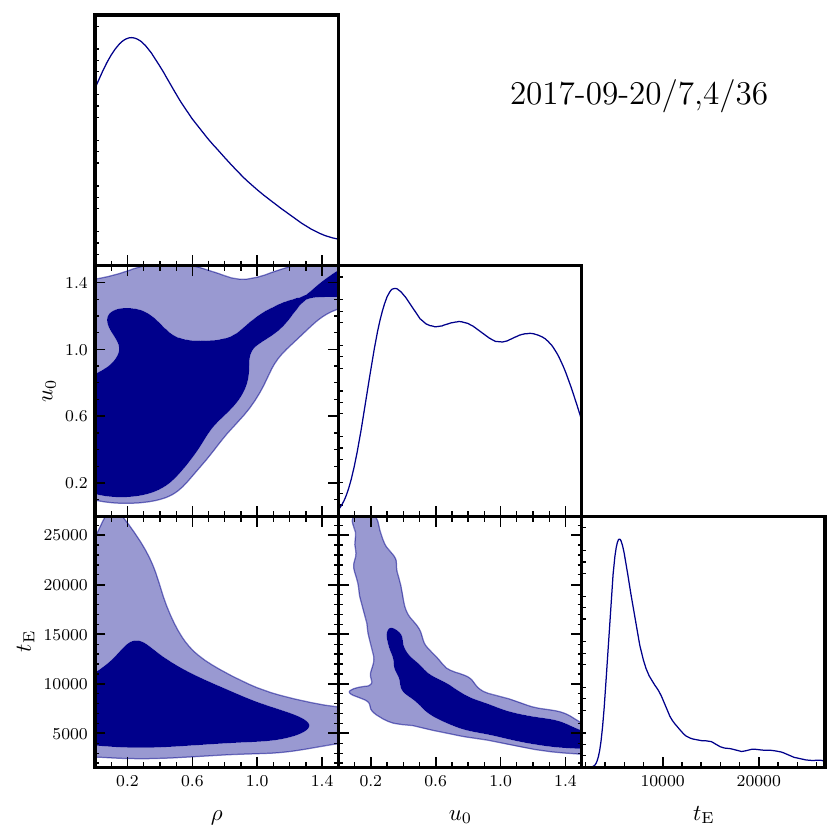}
\includegraphics[width=0.3\textwidth]{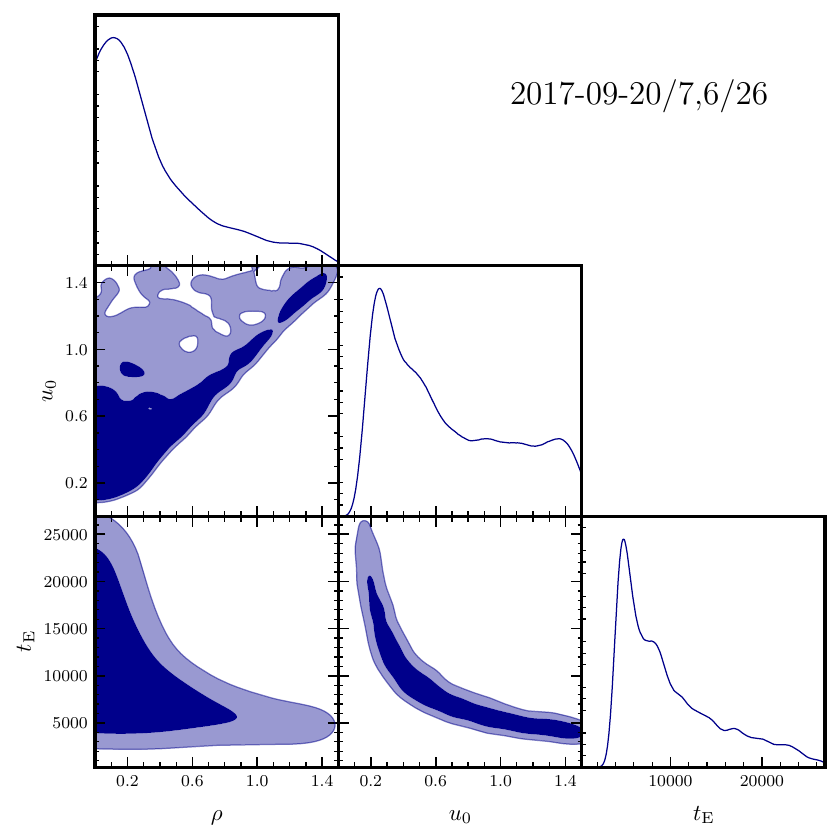}
\includegraphics[width=0.3\textwidth]{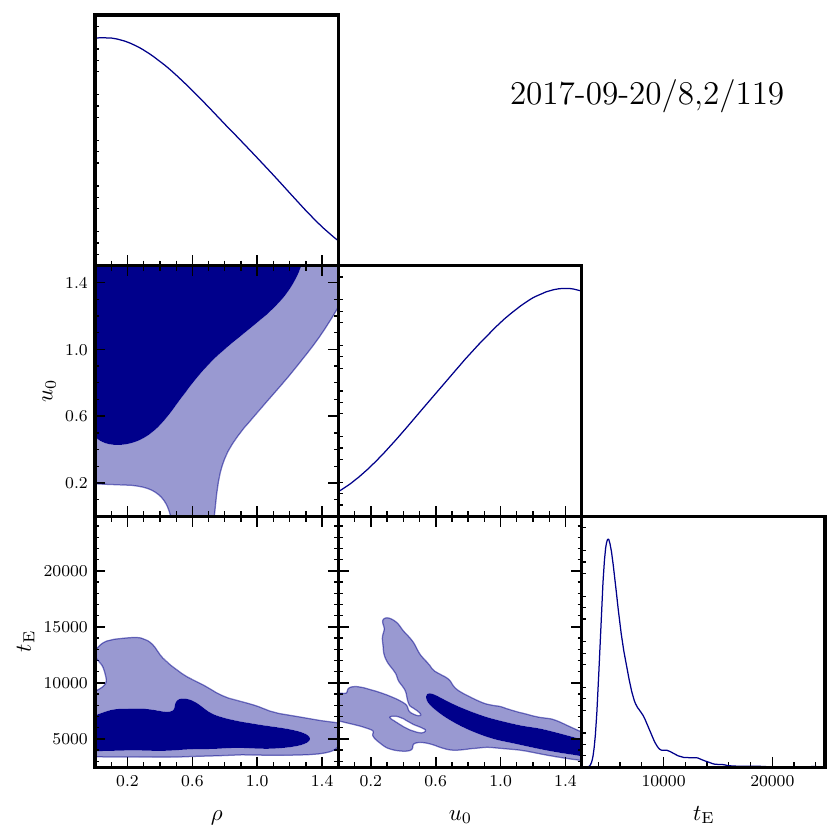}
\caption{Posteriors of the PLFS parameters, $\bm{\theta}=(\rho, u_0, t_{\rm E}, \hat{t})$ for the events in the candidate catalog. The first five panels show the posteriors for events of 2014-11-24, and the rest for 2017-09-20 data.}
\label{fig:lc-posteriors}
\end{figure*}
In this appendix, we present the posterior distribution of the model parameters for each selected light-curve in Section~\ref{ssec:event-selection}. We only shows the 1D/2D marginalized posterior for $\rho, u_0, t_{\rm E}$, and a derived parameter $\hat{t}$. We found that the parameter degeneracy is strong for $\rho, u_0, t_{\rm E}$ for most of the events, and $\hat{t}$ has bounded posterior distribution with less degeneracy than others. This motivates the use of $\hat{t}$ as a nominal time scale parameter of each event.

\section{Lightcurve likelihood for hierarchical Bayesian inference}
\label{apdx:lc-like-mc-integral}
In this appendix, we explain how to perform the integral in the denominator of in Eq.~(\ref{eq:lc-like}).
Here, we only consider one light curve data $\bm{d}_j^n$, but the same applies to other light curve data to construct the likelihood for a set of observed light curves.

As is discussed in \citet{Mandel.Gair.2019}, the likelihood can be rewritten in terms of the posterior of the PLFS parameters for a given light curve data,
\begin{align}
    \int\dd d\int\dd\bm{\theta} P(\bm{d}_j^n | \bm{\theta}) 
    \frac{\dd\Gamma}{\dd d\dd\bm{\theta}}(\bm{\lambda}) = 
    P(\bm{d}_j^n)
    \int\dd d\int\dd\bm{\theta} \frac{P(\bm{\theta} | \bm{d}_j^n) }{\Pi(\bm{\theta})}
    \frac{\dd\Gamma}{\dd d\dd\bm{\theta}}(\bm{\lambda})
\end{align}
where $P(\bm{d}_j^n)$ is the data evidence and $\Pi(\bm{\theta})$ is the parameter prior used in the analysis (see Table~\ref{tab:plfs-prior} for our PLFS case).
We used flat prior for the PLFS parameters for the parameter inference, and therefore we can effectively neglect the effect of the prior as far as we keep the integral variable to be $\bm{\theta}=\{t_{\rm E}, u_0, \rho\}$.
We also omit $P(\bm{d}_j^n)$, because it only gives constant factor and does not depends on the population parameter $\bm{\lambda}$.

We wish to substitute the differential eventrate introduced in Eq.~(\ref{eq:differential-eventrate}) into the above expression. However, 
the differential eventrate in Appendix~\ref{apdx:eventrate-model} is a function of $\bm{\theta}'=\{\hat{t}, \theta, R_{\rm s}\}$, so we need to account for the Jacobian, which is
\begin{align}
    \frac{\dd\Gamma}{\dd d\dd\bm{\theta}} = 
    \frac{\partial\bm{\theta}'}{\partial\bm{\theta}}
    \frac{\dd\Gamma}{\dd d\dd\bm{\theta}'}
    \equiv
    J\frac{\dd\Gamma}{\dd d\dd\bm{\theta}'}
\end{align}
The variable $\bm{\theta}'$ can be written in terms of $\bm{\theta}$ as
\begin{align}
    \begin{cases}
        \hat{t}   &= 2t_{\rm E}\sqrt{u_{\rm T}^2(\rho) - u_0^2} \\
        R_{\rm s} &= \rho d_{\rm s}\theta_{\rm E} \\
        \theta    &= \arccos\left(\frac{u_0}{u_{\rm T}}\right)
    \end{cases}.
    \label{eq:change-of-plfs-param}
\end{align}
We can easily find that the Jacobian matrix is a triangle matrix (after appropriate permutation of variables), and therefore the determinant of the Jacobian matrix becomes as simple as
\begin{align}
    |J| = 
    \left|\frac{\partial \hat{t}}{\partial t_{\rm E}}\right|
    \left|\frac{\partial R_{\rm s}}{\partial \rho}\right|
    \left|\frac{\partial \theta}{\partial u_0}\right| = 
    2d_{\rm s}\theta_{\rm E}
\end{align}
Including the Jacobin determinant appropriately, we obtain the numerator of the light-curve likelihood as
\begin{align}
    \int\dd d\int\dd\bm{\theta} 
    P(\bm{\theta} | \bm{d}_j^n)
    |J| \frac{\dd\Gamma}{\dd d\dd\bm{\theta}'}(\bm{\lambda}) 
    = 
    \sum_h
    \int\dd d 
    \frac{\rho_{{\rm DM},h}(d)v_{{\rm c},h}^2(d)}{M_{\rm PBH}}
    \frac{d_{\rm s}R_{\rm E}}{d}
    \int\dd\bm{\theta} 
    P(\bm{\theta}|\bm{d}_j^n)
    f_{\rm s}(R_{\rm s})
    \left(\frac{v_{\rm r}}{v_{{\rm c},h}}\right)^4
    \exp\left[
    -\left(\frac{v_{\rm r}}{v_{{\rm c},h}}\right)^2
    \right]
\end{align}
where $R_{\rm s}$ and $v_{\rm r}$ are now defined as functions of $\bm{\theta}$;
$R_{\rm s}(\bm{\theta})$ depends on $\rho$ as in Eq.~(\ref{eq:change-of-plfs-param}), and $v_{\rm r}(\bm{\theta}) = R_{\rm E}/t_{\rm E}$, and we omit the dependence on $d$ and $M_{\rm PBH}$ for simplicity of notation now.
We evaluate the above $\bm{\theta}$ integral by Monte-Carlo integration using the samples of PLFS parameter set $\bm{\theta}$ obtained from the light curve fitting in Section~\ref{ssec:candidate-secure-catalog}.
\begin{align}
    \sum_h
    \int\dd d 
    \frac{\rho_{{\rm DM},h}(d)v_{{\rm c},h}^2(d)}{M_{\rm PBH}}
    \frac{d_{\rm s}R_{\rm E}}{d}
    \sum_{q \in \text{MC chain}}
    f_{\rm s}(R_{\rm s}(\bm{\theta}_q))
    \left(\frac{v_{\rm r}(\bm{\theta}_q)}{v_{{\rm c},h}}\right)^4
    \exp\left[
    -\left(\frac{v_{\rm r}(\bm{\theta}_q)}{v_{{\rm c},h}}\right)^2
    \right]
\end{align}
Note that we do not have $P(\bm{\theta}|\bm{d}_j^n)$ in the above equation because the samples to evaluate the integral follow the probability density already. 
We appropriately include the weights of the MC samples, because our samples are obtained by a nested sampler \code{MultiNest} \cite{Feroz.Bridges.2009}, which provides the weights of the samples.

\section{Testing some extended mass function of PBH}
\label{apdx:test-model}
\begin{figure*}
    \includegraphics[width=0.45\textwidth]{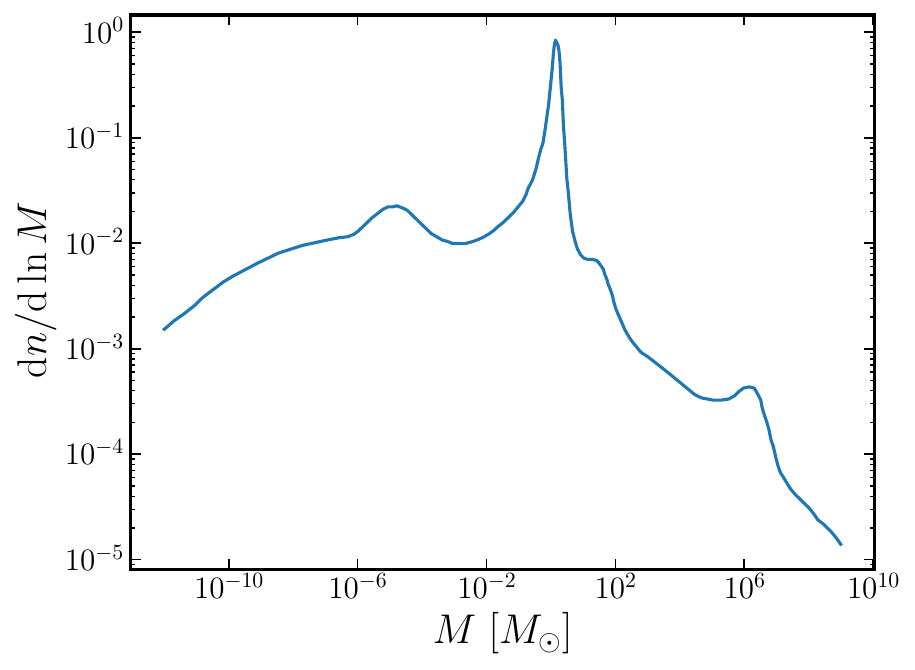}
    \includegraphics[width=0.45\textwidth]{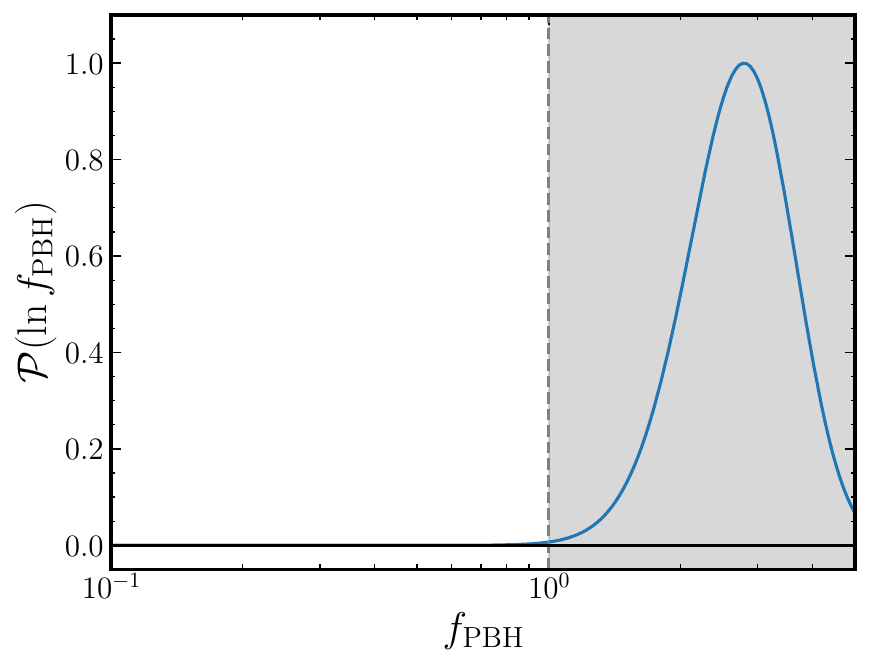}
    \caption{{\it Left}: The shape of the extended mass function proposed by \citet{Carr.Kuhnel.2023}. {\it Right}: The posterior distribution of the amplitude of the mass function, obtained by using the secure events of this work.}
    \label{fig:model-carr}
\end{figure*}
\begin{figure*}
    \includegraphics[width=0.4\textwidth]{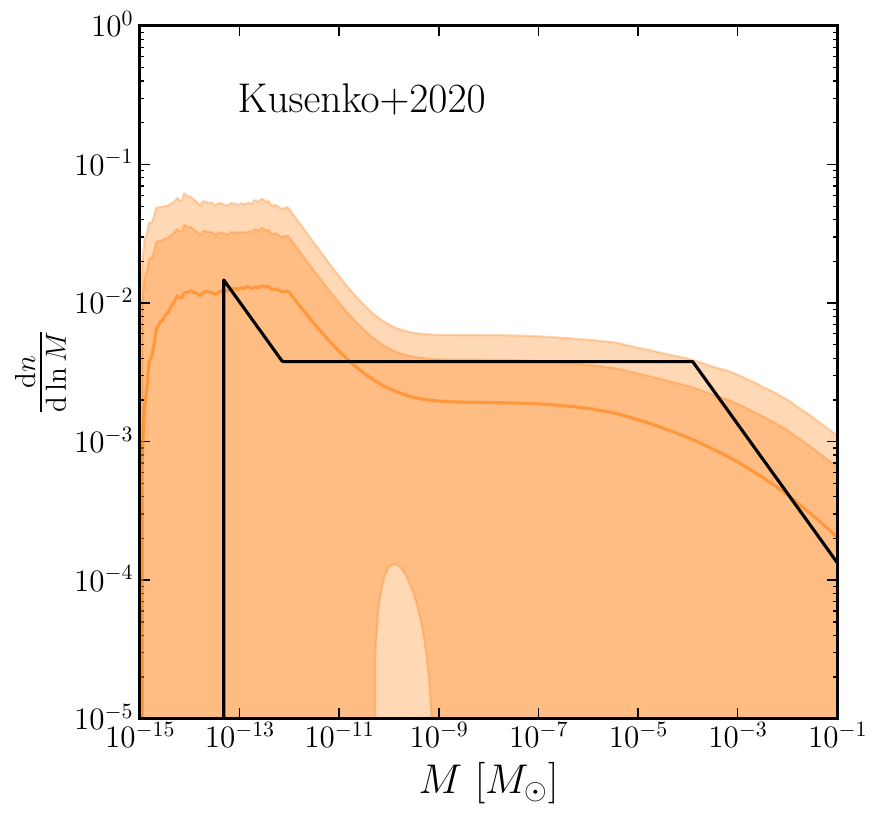}
    \includegraphics[width=0.4\textwidth]{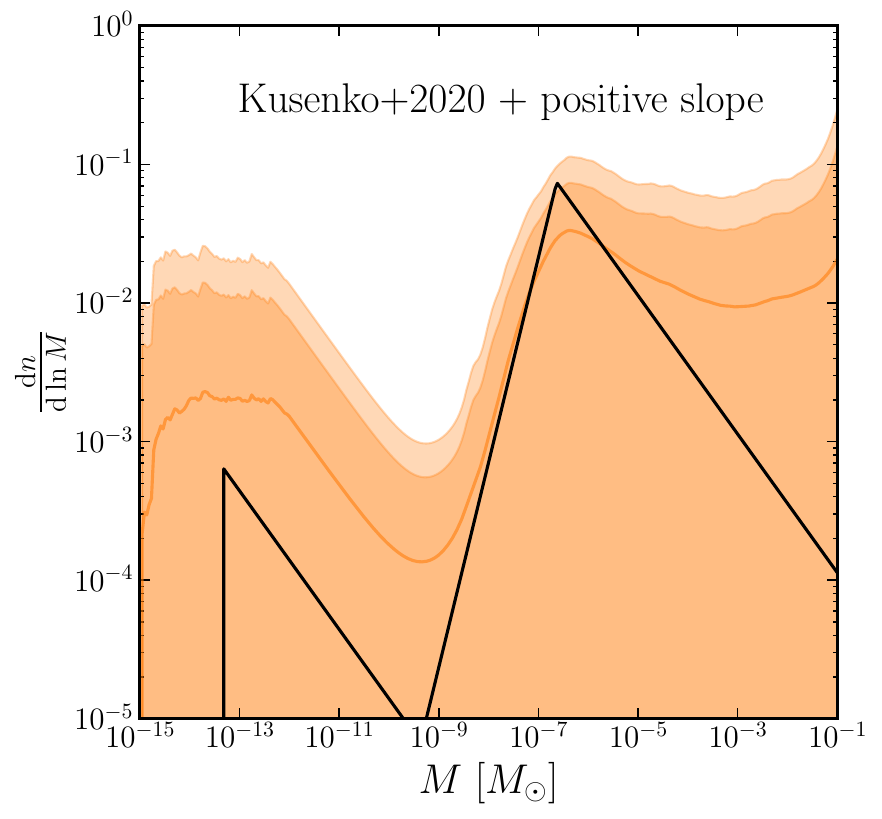}
    \caption{The posterior distribution of the extended mass 
    function proposed in \citet{Kusenko.Vitagliano.2020}. {\it Left}: 
    The light- and dark-shaded region is the 68\% and 95\% credible 
    distribution of the extended mass function obtained using the 
    posterior samples of the model parameters of Eq.~(\ref{eq:kusenko-model}). 
    The reason why the mass function is smooth instead of step like is the 
    step-like mass functions are averaged over for various model parameters 
    in the posterior. The black lines are the the mass function obtained by 
    using the best-fit model in the posterior samples. {\it Right}: The 
    similar figure as the left, but the model with an additional free 
    parameter allowing positive slope in the intermediate mass scale.}
    \label{fig:model-kusenko}
\end{figure*}

In this appendix, we present some example tests of 
the PBH mass function that have more extended shape in mass function.
As the first test example, we consider the model by 
\citet{Carr.Kuhnel.2023}. 
The shape of the mass function is motivated by some candidate of 
positive signals of the PBH from various observations, 
and its mass function shape is shown in 
Fig.~\ref{fig:model-carr}.

To test this model, we introduce a free parameter, the overall 
amplitude of abundance $f_{\rm PBH}$, while fixing the spectrum shape 
of the mass function. 
We infer the amplitude using the secure events 
by following the method in 
Section~\ref{sec:likelihood-model}. 
The resultant posterior distribution of $f_{\rm PBH}$ is shown in Fig.~\ref{fig:model-carr}, where we see that the HSC result prefers $f_{\rm PBH}>1$, which indicates that the mass function has too small amplitude at the mass scale to explain the HSC secure events.

Next, we consider the mass function predicted by 
\citet{Kusenko.Vitagliano.2020}. The mass function has the 
following functional form
\begin{align}
    \frac{{\rm d}n}{{\rm d}\ln M} \propto 
    \begin{cases}
        0                   & (M<M_{\min}) \\
        (M/M_{\min})^{-1/2}   & (M_{\min} < M < M_1) \\
        (M_1/M_{\min})^{-1/2} & (M_1 < M < M_2) \\
        (M_1/M_{\min})^{-1/2} (M/M_2)^{-1/2} & (M_2 < M) 
    \end{cases}
    \label{eq:kusenko-model}
\end{align}
with $M_1$ and $M_2$ free parameters associated with the beginning and the end time of the intermediate matter dominated era in the early Universe.

Fig.~\ref{fig:model-kusenko} shows the constraint on the mass function.
We notice that although we have a non-zero events in the PBH hypothesis,
the overall amplitude is strongly suppressed. 
This is because, even in the PBH hypothesis, the PBH mass function is 
allowed to be abundant only at the mass scale of $10^{-7}M_\odot$ 
and the smaller abundance is preferred at the lower mass scale. 
For this model, the mass function is extended broadly over 
the various mass scale, and its shape is fixed in the mass scale of HSC sensitivity, and therefore these effects compete, 
and as a consequence the smaller total abundance is preferred.

Although it is not motivated by the theory, we introduce an additional 
free parameter in this model to understand how the constraint changes
when we change the shape of the mass function in the HSC sensitivity scales,
\begin{align}
    \frac{{\rm d}n}{{\rm d}\ln M} \propto 
    \begin{cases}
        0                              & (M<M_{\min}) \\
        (M/M_{\min})^{-1/2}            & (M_{\min} < M < M_1) \\
        (M_1/M_{\min})^{-1/2}(M/M_1)^p & (M_1 < M < M_2) \\
        (M_1/M_{\min})^{-1/2}(M_2/M_1)^p (M/M_2)^{-1/2} & (M_2 < M) 
    \end{cases}
    \label{eq:kusenko-model-p}
\end{align}
where we allow a positive slope in the mass function for $M_1 < M < M_2$. 
The constraint on this mass function is shown in Fig.~\ref{fig:model-kusenko}.
We see that indeed the positive slope is preferred, which is because the 
abundant PBH is only allowed at the mass scale $M\sim 10^{-7}M_\odot$ and lower abundance is preferred at the lower mass scale, leading highest peak at $M\sim 10^{-7}M_\odot$.

\end{document}